\documentclass[a4paper,twocolumn,aps,prb,floatfix,showpacs,preprintnumbers]{revtex4}
\usepackage[latin9]{inputenc}
\usepackage{amsmath}
\usepackage{amssymb}
\usepackage{graphicx}
\usepackage{esint}

\makeatletter

\@ifundefined{textcolor}{}
{%
 \definecolor{BLACK}{gray}{0}
 \definecolor{WHITE}{gray}{1}
 \definecolor{RED}{rgb}{1,0,0}
 \definecolor{GREEN}{rgb}{0,1,0}
 \definecolor{BLUE}{rgb}{0,0,1}
 \definecolor{CYAN}{cmyk}{1,0,0,0}
 \definecolor{MAGENTA}{cmyk}{0,1,0,0}
 \definecolor{YELLOW}{cmyk}{0,0,1,0}
}

\@ifundefined{textcolor}{}{%
 \definecolor{BLACK}{gray}{0}
 \definecolor{WHITE}{gray}{1}
 \definecolor{RED}{rgb}{1,0,0}
 \definecolor{GREEN}{rgb}{0,1,0}
 \definecolor{BLUE}{rgb}{0,0,1}
 \definecolor{CYAN}{cmyk}{1,0,0,0}
 \definecolor{MAGENTA}{cmyk}{0,1,0,0}
 \definecolor{YELLOW}{cmyk}{0,0,1,0}
}

\makeatother

\begin{document}

\title{Effect of charged line defects on conductivity in graphene: numerical
Kubo and analytical Boltzmann approaches }

\author{T. M. Radchenko,$^{1}$ A. A. Shylau,$^{2}$ and I. V. Zozoulenko$^{3}$}

\affiliation{$\mathit{\mathrm{\textrm{\ensuremath{^{1}}}}}$Department of Solid
State Theory, Institute for Metal Physics, NASU, 36 Acad. Vernadsky
Blvd., 03680 Kyiv, Ukraine}

\affiliation{$\mathit{\mathrm{\textrm{\ensuremath{^{2}}}}}$Department of Micro
and Nanotechnology, DTU Nanotech, Technical University of Denmark,
DK-2800 Kongens Lyngby, Denmark}

\affiliation{$\mathit{\mathrm{\textrm{\ensuremath{^{3}}}}}$Organic Electronics,
Department of Science and Technology (ITN), Linköping University,
60174 Norrköping, Sweden}

\author{Aires Ferreira$^{1}$}

\affiliation{$\mathit{\mathrm{\textrm{\ensuremath{^{1}}}}}$Graphene Research
Center and Department of Physics, National University of Singapore,
2 Science Drive 3, Singapore 117542}

\date{\today }
\begin{abstract}
Charge carrier transport in single-layer graphene with one-dimensional
charged defects is studied theoretically. Extended charged defects,
considered an important factor for mobility degradation in chemically-vapor-deposited
graphene, are described by a self-consistent Thomas--Fermi potential.
A numerical study of electronic transport is performed by means of
a time-dependent real-space Kubo approach in honeycomb lattices containing
millions of carbon atoms, capturing the linear response of realistic
size systems in the highly disordered regime. Our numerical calculations
are complemented with a kinetic transport theory describing charge
transport in the weak scattering limit. The semiclassical transport
lifetimes are obtained by computing scattered amplitudes within the
second Born approximation. The transport electron--hole asymmetry
found in the semiclassical approach is consistent with the Kubo calculations.
In the strong scattering regime, the conductivity is found to be a
sublinear function of electronic density and weakly dependent on the
Thomas--Fermi screening wavelength. We attribute this atypical behavior
to the extended nature \textit{\emph{of one-dimensional}} charged
defects. Our results are consistent with recent experimental reports.\textbf{
} 
\end{abstract}


\maketitle

\section{Introduction\label{sec:Introduction}}

The isolation of graphene---the queen of two-dimensional materials
due to its remarkable physical properties---by the exfoliation method
has triggered intensive studies of its fundamental properties and
has opened horizons for future technologies.\cite{Novoselov2004,Kats2007,Castro Netto review}
Since that time, various methods of graphene growth have been explored
in order to make the fabrication process scalable; a prerequisite
for developing graphene-based devices and technologies.\cite{Geim-Novos2007}
Nowadays, several techniques are capable of producing high-quality,
large-scale graphene. These include epitaxial graphene growth on SiC,\cite{de Heer2007}
and chemical vapor deposition (CVD) of graphene on transition metal
surfaces.\cite{Kim2009} The advantages of the latter method lie in
its low cost, possibility to grow large graphene sheets (tens of inches),
and ease of its transfer into other substrates.\cite{Bae2010} Currently,
there is a strong motivation for exploring electronic and transport
properties of CVD grown graphene because it represents one of the
most promising materials for flexible and transparent electronics.

The studies of the transport properties of graphene are often focused
on the fundamental question: what limits a charge carrier mobility
in it? As far as CVD-grown graphene is concerned, it is believed that
its transport properties are strongly affected by the presence of
charged line defects.\cite{FerreiraEPL} Usually, the growth of graphene
by the CVD-method requires to use metal surfaces with hexagonal symmetry,
such as the (111) surface of cubic or the (0001) surface of hexagonal
crystals.\cite{Krasheninnikov2011} The mismatch between the metal-substrate
and graphene causes the strains in the latter, reconstructs the chemical
bonds between the carbon atoms and results in formation of two-dimensional
(2D) domains of different crystal orientations separated by one-dimensional
defects.\cite{Krasheninnikov2011,Jeong2008,Yazyev2010,Malola2010}
The nucleation of the graphene phase takes place simultaneously at
different places, which leads to the formation of independent 2D domains
matching corresponding grains in the substrate. A line defect appears
when two graphene grains with different orientations coalesce; the
stronger the interaction between graphene and the substrate, the more
energetically preferable the formation of line defects is. These line
defects accommodate localized states trapping the electrons, originating
lines of immobile charges that scatter the Dirac fermions in graphene.

It is well established that the presence of grains and grain boundaries
in three-dimensional polycrystalline materials can strongly affect
their electronic and transport properties. Hence, in principle, the
role of such structures in 2D materials, such as graphene, can be
even more important because even a single line defect can divide and
disrupt the crystal.\cite{Krasheninnikov2011} A series of recent
control experiments\cite{Song2012,Li2009} strongly indicate that
line defects are responsible for lower carrier mobility in CVD-grown
graphene in comparison to the exfoliated samples.\cite{Morozov2008,Bolotin2008,Du2008}
We note in passing that one-dimensional (1D) defects have been observed
not only in experimental studies on CVD growth of graphene films,
for instance, on Cu,\cite{Li2009} Ni,\cite{Lahiri2010} Ir,\cite{Coraux2008}
but also in single graphene layer after electron irradiation\cite{Hashimoto2004}
and in highly oriented pyrolytic graphite surface.\cite{Cervenka2009}
Possible applications include: valley filtering based on scattering
off line defects,\cite{Gunlycke2011} ferromagnetic ordering in line
defects,\cite{Okada2011,Kou2011} enhancement of electron transport\cite{Kindermann2010}
or chemical reactivity\cite{Botello2011} due to induced extra conducting
channels and localized states along the line, quantum channels controlled
by tuning of the gate voltage embedded below the line defect,\cite{SongPRB2012}
and correlated magnetic states in extended defects.\cite{Alexandre2012}

Several\textbf{ }theoretical studies have been recently reported addressing
transport properties of graphene with a single graphene boundary\cite{YazyevNature,Liwei,Peres}
or polycrystalline graphene with many domain boundaries.\cite{Tuan2013}\textbf{
}On the other hand, much less attention has been paid to the effect
of charge accumulation at these boundaries due to self-doping. Transport
properties of graphene with 1D \textit{charged} defects has been studied
in Ref.~\onlinecite{FerreiraEPL} using the Boltzmann approach
within the first Born approximation. It has been demonstrated previously
that such approximation is not always applicable for the description
of electron transport in graphene even at finite (non-zero) electronic
densities.\cite{Klos2010,Ferreira,HengyiXu2011,Radchenko} In the
present work we investigate the impact of extended charged defects
in the transport properties of graphene by an exact numerical approach
based on the time-dependent real-space quantum Kubo method\cite{Roche_SSC,Roche1997,Triozon2002,Triozon2004,Markussen,Ishii,Yuan10,Ferreira,Leconte11,Trambly,Lherbier08100,Lherbier08101,Lherbier11,Radchenko}
which is especially suited to treat large graphene systems with dimensions
approaching realistic systems containing millions of atoms. Our numerical
calculations are complemented with a semi-classical treatment going
beyond the first Born approximation, describing the transport properties
in the weak scattering regime.

The paper is organized as follows. The numerical models (tight-binding
approximation and Kubo approach) and obtained results are presented
in Sec.~\ref{sec:TB_Kubo_RealSpace}. In Sec.~\ref{sec:Boltzmann_Approach}
we study the impact of extended charged defects within kinetic transport
theory. Here, the general expression for the scattering amplitude
for massless fermions within the second Born approximation is derived
and used to obtain the semiclassical conductivity and the transport
electron--hole asymmetry. The approaches in Secs.~\ref{sec:TB_Kubo_RealSpace}~and~\ref{sec:Boltzmann_Approach}
provide information about transport dominated by 1D charged defects
in distinct regimes. Section~\ref{sec:Conclusions} presents the
conclusions of our work. Details of numerical calculations and analytic
derivations are given in the Appendixes.

\section{Tight-binding model and time-dependent real-space Kubo--Greenwood
formalism\label{sec:TB_Kubo_RealSpace}}

In this section, we introduce to the basis of the tight-binding approximation
as well as the Kubo--Greenwood approach and also present numerical
results obtained within the framework of these models.

\subsection{Basics\label{sub:sub_Basics}}

To model electron dynamics in graphene, we use a standard $p$-orbital
nearest neighbor tight-binding Hamiltonian defined on a honeycomb
lattice\cite{Castro Netto review,Peres_review,DasSarma_review}

\begin{equation}
\hat{H}=-u\sum_{i,i^{\prime}}c_{i}^{\dagger}c_{i^{\prime}}+\sum_{i}V_{i}c_{i}^{\dagger}c_{i},\label{Eq_Hamiltonian}
\end{equation}
where $c_{i}^{\dagger}$ and $c_{i}$ are the standard creation and
annihilation operators acting on a quasiparticle on the site $i$.
The summation over $i$ runs over the entire graphene lattice, while
$i^{\prime}$ is restricted to the sites next to $i$; $u=2.7$~eV
is the hopping integral for the neighboring C atoms $i$ and $i^{\prime}$
with distance $a=0.142$~nm between them, and $V_{i}$ is the on-site
potential describing impurity (defect) scattering.

\begin{figure}
\centering{}\includegraphics[width=1\columnwidth]{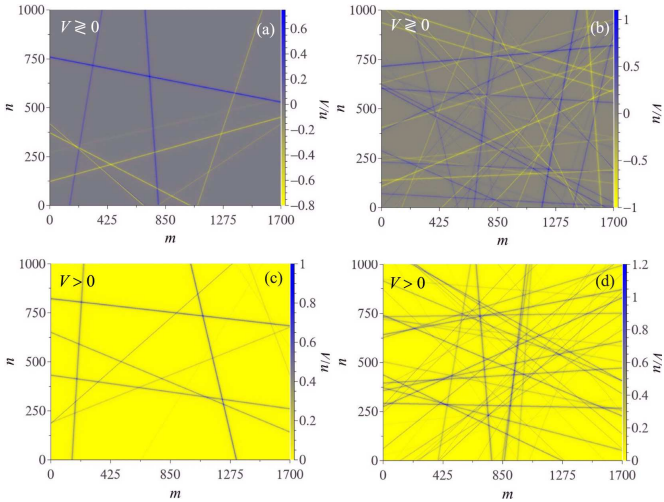} \caption{\label{Fig_Potential}(Color online) Effective symmetric {[}(a),(b){]}
and asymmetric {[}(c),(d){]} potentials describing one of the possible
configurations of 10 (left) and 50 (right) line defects in graphene
sheet of the size $m\times n=1700\times1000$ sites corresponding
to $210\times210$ nm. Maximal potential height $\triangle=0.25u$. }
\end{figure}

Since line defects can be thought as lines of reconstructed point
defects,\cite{Jeong2008,Krasheninnikov2011,Yazyev2010,Malola2010}
we model a 1D defect as point defects oriented along a fixed direction
(corresponding to the line direction) in the honeycomb lattice. The
electronic effective potential for a charged line within the Thomas--Fermi
approximation was first obtained in Ref.~\onlinecite{FerreiraEPL}
(see also Appendix A); if there are $N_{\text{lines}}$ such charged
lines in a graphene lattice, the effective scattering potential reads
as 
\begin{align}
V_{i}= & \sum_{j=1}^{N_{\text{lines}}}U_{j}[-\text{cos}(q_{\mathrm{\textrm{TF}}}x_{ij})\text{Ci}(q_{\textrm{TF}}x_{ij})+\notag\\
 & +\text{sin}(q_{\textrm{TF}}x_{ij})(\pi/2-\text{Si}(q_{\textrm{TF}}x_{ij}))],\label{Eq_TF potential}
\end{align}
where $U_{j}$ is a potential height, $x_{ij}$ is a distance between
the site $i$ and the $j$-th line, $q_{\textrm{TF}}=e^{2}k_{F}/(\pi\varepsilon_{0}\varepsilon_{\textrm{r}}\hbar v_{F})$
is the Thomas--Fermi wavevector defined by the electron Fermi velocity
$v_{F}=3ua/(2\hbar)$ and the Fermi momentum $k_{F}=\sqrt{\pi|n_{e}|}$
(related to the electronic carrier density $n_{e}$ controlled applying
the back-gate voltage). Here, $-e<0$ denotes the electron charge.
The Thomas--Fermi wavevector is also commonly expressed as a function
of graphene's structure constant $\alpha_{\textrm{g}}\equiv e^{2}k_{F}/(4\pi\varepsilon_{0}\varepsilon_{\textrm{r}}\hbar v_{F})$
according to $q_{\textrm{TF}}=4\alpha_{\textrm{g}}k_{F}$. We consider
two cases: symmetric, $V\gtrless0$, and asymmetric, $V>0$, potentials,
where $U_{j}$ are chosen randomly in the ranges $\left[-\triangle,\triangle\right]$
and $\left[0,\triangle\right]$, respectively, with $\triangle$ being
the maximal potential height. In order to simplify numerical calculations,
we fit the potential (\ref{Eq_TF potential}) by the Lorentzian function
\begin{equation}
V_{i}=\sum_{j=1}^{N_{\text{lines}}}U_{j}(A/(B+Cx_{ij}^{2}))\label{eq:Lor}
\end{equation}
as described in Appendix~A. A typical shape of the effective potential
for both symmetric and asymmetric cases is illustrated in Fig.~\ref{Fig_Potential}.

\subsection{Time-dependent real-space Kubo method}

To calculate numerically the dc conductivity $\sigma$ of graphene
sheets with 1D charged defects, the real-space order-$N$ numerical
implementation within the Kubo--Greenwood formalism is employed, where
$\sigma$ is extracted from the temporal dynamics of a wave packet
governed by the time-dependent Schrödinger equation.\cite{Roche_SSC,Roche1997,Trambly,Triozon2002,Triozon2004,Yuan10,Markussen,Ishii,Leconte11,Lherbier08100,Lherbier08101,Lherbier11,Tuan2013}
This is a computationally efficient method scaling with a number of
atoms in the system $N$, and thus allowing treating very large graphene
sheets containing many millions of C atoms.

A central quantity in the Kubo--Greenwood approach is the mean quadratic
spreading of the wave packet along the $x$-direction at the energy
$E$, $\Delta\hat{X}^{2}(E,t)=\bigl\langle\hat{(X}(t)-\hat{X}(0))^{2}\bigr\rangle$,
where $\hat{X}(t)=\hat{U}^{\dagger}(t)\hat{X}\hat{U}(t)$ is the position
operator in the Heisenberg representation, and $\hat{U}(t)=e^{-i\hat{H}t/\hbar}$
is the time-evolution operator. Starting from the Kubo--Greenwood
formula for the dc conductivity\cite{Madelung}

\begin{equation}
\sigma=\frac{2\pi\hbar e^{2}}{\Omega}\text{Tr}[\hat{v}_{x}\delta(E-\hat{H})\hat{v}_{x}\delta(E-\hat{H})],\label{Eq_Kubo-Greenwood}
\end{equation}
where $\hat{v}_{x}$ is the $x$-component of the velocity operator,
$E$ is the Fermi energy, $\Omega$ is the area of the graphene sheet,
and factor 2 accounts for the spin degeneracy, the conductivity can
then be expressed as the Einstein relation, 
\begin{equation}
\sigma\equiv\sigma_{xx}=e^{2}\tilde{\rho}(E)\lim_{t\rightarrow\infty}D(E,t),\label{Eq_sigma(t)}
\end{equation}
where $\tilde{\rho}(E)=\rho/\Omega=\textrm{Tr}[\delta(E-\hat{H})]/\Omega$
is the density of sates (DOS) per unit area (per spin), and the time-dependent
diffusion coefficient $D(E,t)$ relates to $\Delta\hat{X}^{2}(E,t)$
according to 
\begin{align}
D(E,t)= & \frac{\bigl\langle\Delta\hat{X}^{2}(E,t)\bigr\rangle}{t}\notag\\
= & \frac{1}{t}\frac{\text{Tr}[\hat{(X}_{H}(t)-\hat{X}(0))^{2}\delta(E-\hat{H})]}{\text{Tr}[\delta(E-\hat{H})]}.\label{Eq_Diffusion}
\end{align}

It should be noted that in the present study we are interested in
the diffusive transport regime when the diffusion coefficient reaches
its maximum. Therefore, following Refs.~\onlinecite{Leconte11}
and \onlinecite{Lherbier12}, we replace in Eq.~(\ref{Eq_sigma(t)})
$\lim_{t\rightarrow\infty}D(E,t)\rightarrow D_{\max}(E),$ such that
the dc conductivity is defined as 
\begin{equation}
\sigma=e^{2}\tilde{\rho}(E)D_{\max}(E).\label{Eq_sigmaMax}
\end{equation}
Note that in most experiments, the conductivity is measured as a function
of electron density $n_{e}$. We calculate the electron density as
$n_{e}(E)\equiv n_{e}=\int_{-\infty}^{E}\tilde{\rho}(E)dE-n_{\text{ions}},$
where $n_{\text{ions}}=3.9\cdot10^{15}$ cm$^{-2}$ is the density
of the positive ions in the graphene lattice compensating the negative
charge of the $p$-electrons {[}note that for the ideal graphene lattice
at the neutrality point $n(E)=0${]}. Combining the calculated $n_{e}(E)$
with $\sigma(E)$ given by Eq.~(\ref{Eq_sigmaMax}) we obtain the
required dependence of the conductivity $\sigma=\sigma(n_{e})$.

\subsection{Numerical results\label{sub:sub_Numerical-results}}

\begin{figure}[b]
\begin{centering}
\includegraphics[width=1\columnwidth]{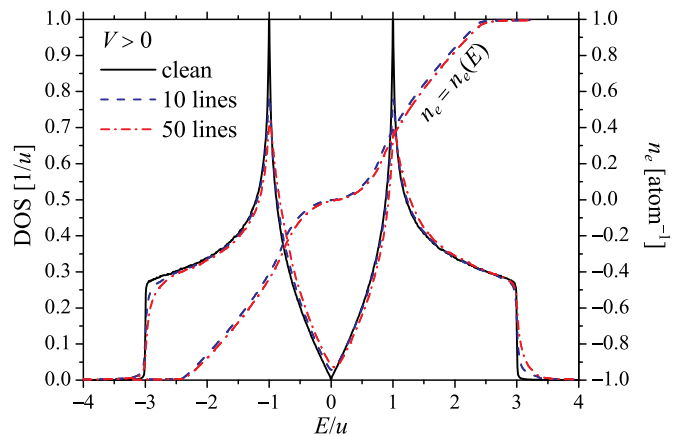} 
\par\end{centering}

\centering{}\caption{\label{Fig_DOS} (Color online) Density of states (DOS) and the relative
charge carrier concentration $n_{e}$ (the number of electrons per
C atom) vs. the energy $E$ for 10 and 50 positively-charged line
defects described by the symmetric potential potential with $\triangle=0.25u$.}
\end{figure}

\begin{figure*}
\begin{centering}
\includegraphics[width=1\textwidth]{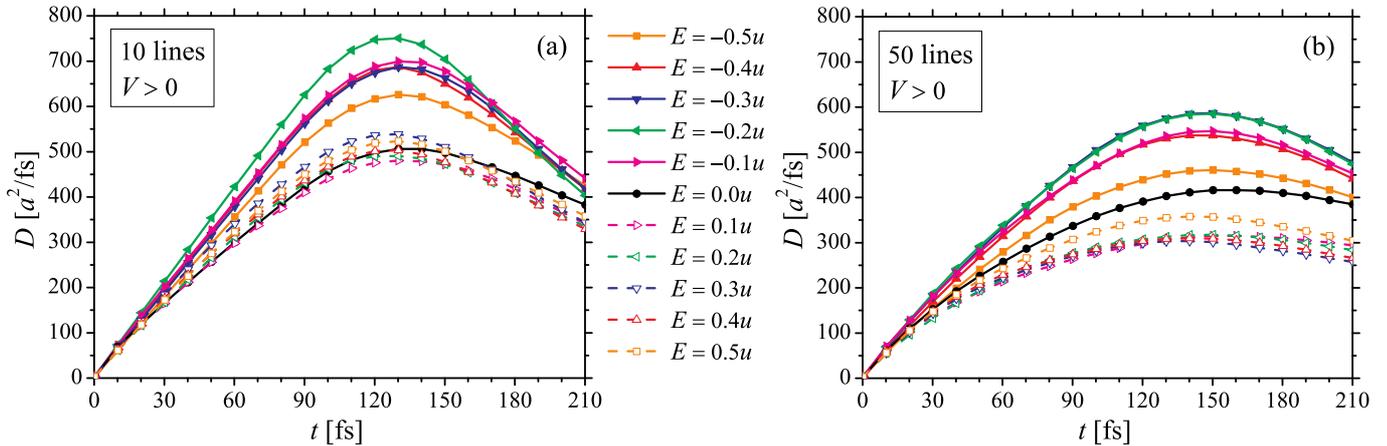} 
\par\end{centering}

\centering{}\caption{\label{Fig_Diffus} (Color online) Time-dependent diffusion coefficient
at different energies for 10 (a) and 50 (b) positively-charged line
defects ($\triangle=0.25u$).}
\end{figure*}

This subsection presents numerical results for the dc conductivity
calculated using the time-dependent real space Kubo--Greenwood formalism
within the tight-binding model. We compute the density dependence
of the conductivity for graphene sheets with 10 and 50 lines in $1700\times1000$
lattice. This approximately corresponds to a relative concentration
of point defects of respectively 1\% and 5\%. We model the potential
due to lines of charges by the Lorentzian function, Eq.~(\ref{eq:Lor}),
where we set $q_{\textrm{TF}}a=0.1$, which corresponds to typical
electron densities $|n_{e}^{\exp}|\sim5\cdot10^{-5}$ atom\texttt{$^{-1}$}
($|n_{e}^{\exp}|\sim2\cdot10^{11}$ cm\texttt{$^{-2}$}), see Fig.~\ref{Fig_Potential}.
It should be noted that $q_{\textrm{TF}}$ is not a constant but weakly
density dependent ($q_{\textrm{TF}}\propto\sqrt{|n_{\textrm{e}}|}$).
Quite remarkably, the obtained results for the conductivity remain
practically unchanged when we use different $q_{\textrm{TF}}$ corresponding
to representative electron densities considered in the present study,
$1\cdot10^{-5}\lesssim|n_{e}|\lesssim5\cdot10^{-5}$ atom\texttt{$^{-1}$}.
In Appendix B we present results of more elaborated self-consistent
calculations where we use the exact shape of the Thomas--Fermi potential
{[}i.e.,~Eq.~(\ref{Eq_TF potential}) instead of Eq.~(\ref{eq:Lor}){]}
and take into account the density dependence of $q_{\textrm{TF}}$.
We found that even for a single charged line embedded in a graphene
sheet, the dependence $\sigma=\sigma(n_{e})$ calculated for the exact
self-consistent (i.e.,~$n_{e}$-dependent) potential (\ref{Eq_TF potential})
exhibits qualitatively and quantitatively the same sublinear behavior
as in the simulations with the fixed Thomas--Fermi wavevector ($q_{\textrm{TF}}a=0.1$)
and with $V_{i}$ given by the Lorentzian function Eq.~(\ref{eq:Lor}).
The same conclusion holds for samples with 10 and 50 lines. Because
of this in what follows we will discuss the results for the case of
the Lorentzian potential at the fixed $q_{\textrm{TF}}$.

\begin{figure*}
\centering{}\includegraphics[width=1\textwidth]{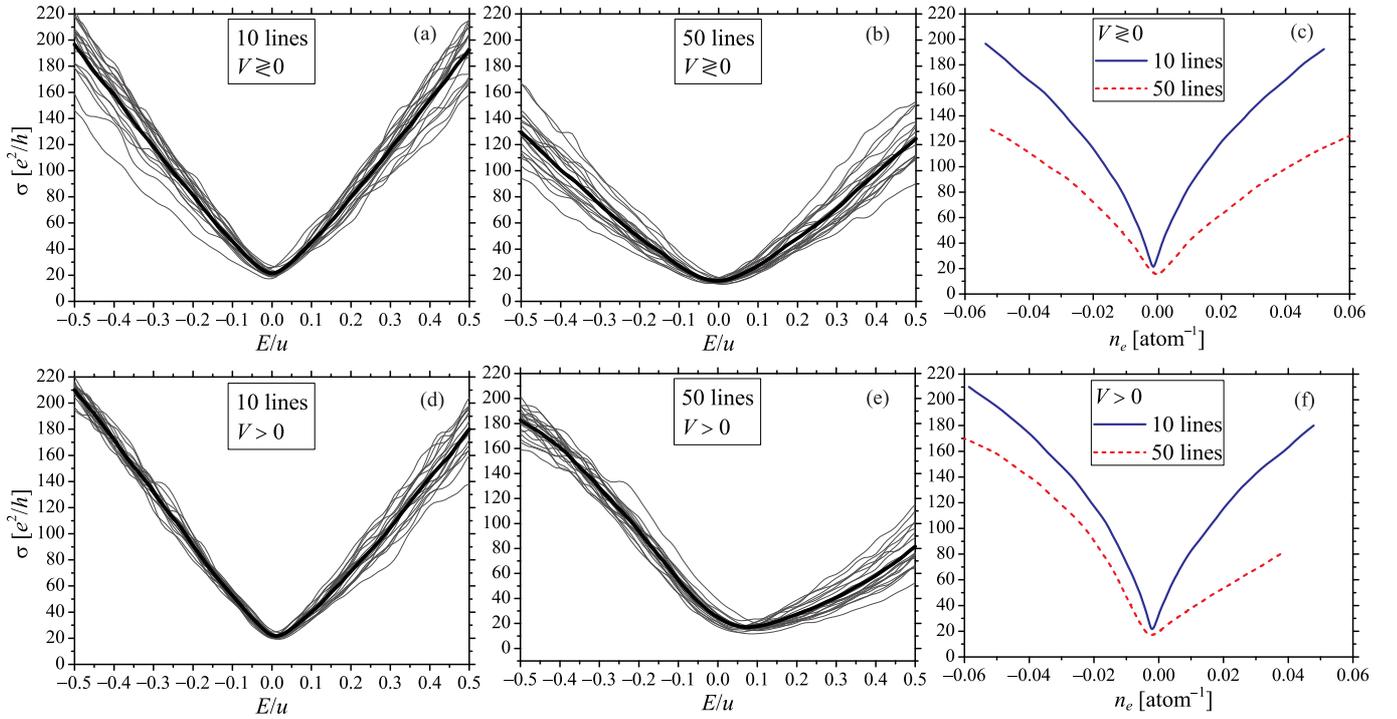} \caption{\label{Fig_Cond_posit-negat_posit} (Color online) Conductivity as
a function of energy $E$ {[}(a),(b),(d),(e){]} and relative electron
density $n$ {[}(c),(f){]} for different configurations of 10 and
50 positively-charged 1D defects ($\triangle=0.25u$). Conductivities
in (c) and (f) are averaged over 20 different realizations in (a),
(b), and (d), (e).}
\end{figure*}

\begin{figure*}
\centering{}\includegraphics[width=0.85\textwidth]{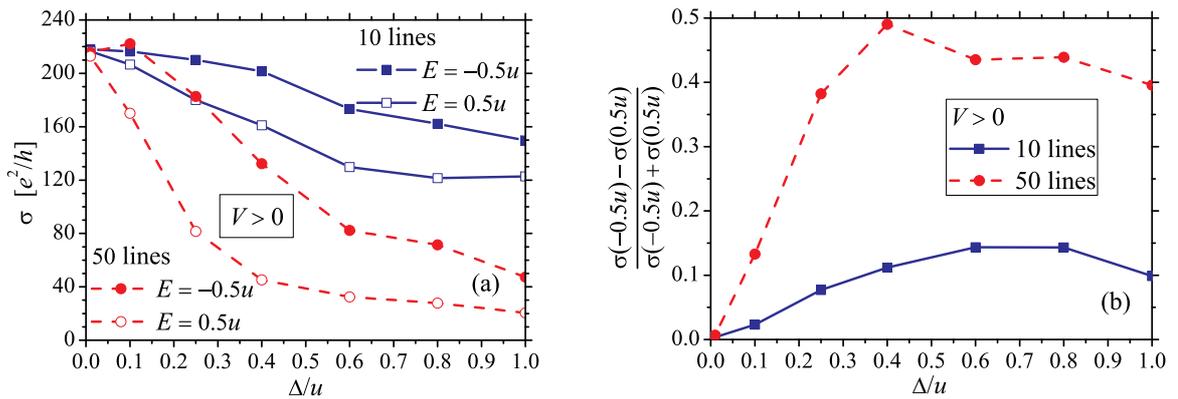} \caption{\label{Fig_asymmetry} (Color online) (a) The conductivity and and
relative (b) values of conductivities for two symmetrical (with respect
to the Dirac point) energies as the functions of positive (asymmetric)
potential $V\sim U\in[0,\triangle]$.}
\end{figure*}

Figure~\ref{Fig_DOS} shows the electron density $n_{e}=n(E)$ and
the DOS in a graphene sheet with different number of charged lines.
The calculated dependencies are very much similar to those for clean
graphene and for graphene with a long-range Gaussian potential.\cite{Yuan10,Radchenko}
(Note that the DOS of graphene with short-range strong scatterers
exhibits an impurity peak in the vicinity of neutrality point.)\cite{Pereira06,peres2006,Yuan10,Ferreira,Radchenko}
For both symmetric (not shown here) and asymmetric potentials, the
DOS does not reach zero at the Dirac point, and the asymmetric potential
(in contrast to the symmetric one) leads to electron--hole asymmetry
in the DOS.

The time dependence of the diffusion coefficient at different energies
for the case of a symmetric potential corresponding to 10 and 50 positively
charged line defects is shown in Fig.~\ref{Fig_Diffus}. {[}Diffusivity
curves for the case of asymmetric potential (not shown here) exhibit
similar behavior.{]} After an initial linear increase corresponding
to the ballistic regime, the diffusion coefficient reaches its maximum
at $t\thickapprox130$ and 150 fs for 10 and 50 lines, respectively.
These values of $D=D_{\max}$ are used to calculate $\sigma$ according
to Eq.~(\ref{Eq_sigmaMax}). For times $t\gtrsim150$, $D(t)$ decreases
due to the localization effects. Similar temporal behavior of the
diffusion coefficient was established earlier for different types
of scatterers in graphene including long-range Gaussian and short-range
potentials.\cite{Leconte11,Lherbier12,Radchenko}

In Fig.~\ref{Fig_Cond_posit-negat_posit} we show the density dependence
of the conductivity of graphene sheets with linear defects for the
cases of symmetric and asymmetric potentials. The obtained dependencies
show following features.

First, the averaged conductivities exhibit a pronounced sublinear
density dependence, see Figs.~\ref{Fig_Cond_posit-negat_posit}(c)
and \ref{Fig_Cond_posit-negat_posit}(f). Our numerical calculations
are consistent with the recent experimental results for the CVD-grown
graphene\cite{Kim2009,Song2012,Tsen2010} that also exhibit sublinear
density dependence. This provides an evidence in support that the
line defects represent the dominant scattering mechanism in CVD-grown
graphene.\cite{FerreiraEPL,Kim2009,Song2012} Note that the calculated
sublinear density dependence for the case of linear defects is quite
different from the case of short- and long-range point scatterers
where the numerical calculations show a density dependence which is
close to linear.\cite{Lewenkopf,Klos2010,Yuan10,Ferreira,Radchenko}

Second, the conductivities of samples with different impurity configurations
exhibit significant variations between each other, see Figs.~\ref{Fig_Cond_posit-negat_posit}(a)--\ref{Fig_Cond_posit-negat_posit}(b),
\ref{Fig_Cond_posit-negat_posit}(d)--\ref{Fig_Cond_posit-negat_posit}(e).
This is in strong contrast to the case of short- and long-range point
scatterers where corresponding conductivities of samples of the same
size and impurity concentrations practically did not show any noticeable
differences for different impurity configurations.\cite{Radchenko}
We attribute this to the fact that in contrast to point defects, the
line defects are characterized not only by their positions, but also
by directions (orientations) and their intersections as well. Such
additional characteristics result in much more possible distributions
of the potential which, in turn, leads to the differences in the conductivity
curves.

Third, for the symmetric potential the conductivity curves are symmetric
with respect to the neutrality point, while the asymmetric one shows
the asymmetry of the conductivity, c.f.,~Figs. \ref{Fig_Cond_posit-negat_posit}(c)
and \ref{Fig_Cond_posit-negat_posit}(f). Such asymmetry between the
holes and electrons have been also reported in many transport calculations
for graphene with point defects, for instance in Refs.~\onlinecite{Lherbier08101,Robinson,Wehling,Leconte11,Ferreira}
and \onlinecite{Radchenko}. For a closer inspection of the effect
of asymmetry we plotted the conductivities for representative energies
$E=\pm0.5u$, Fig.~\ref{Fig_asymmetry}(a) as well as their relative
differences \textbf{$\frac{\sigma(-E)-\sigma(E)}{\sigma(-E)+\sigma(E)}$}
as a function of the potential strength $\Delta$, Fig.~\ref{Fig_asymmetry}(b).
The relative conductivity difference exhibits a linear behavior for
$\Delta\lesssim0.4u$ followed by saturation for larger values of
$\Delta$. A comparison of the obtained numerical results with the
analytic predictions in the weak scattering regime will be given in
what follows.

We conclude this section by noting that conductivity of large CVD-grown
graphene polycrystalline samples with disordered grain boundaries
was calculated by Tuan et al.\cite{Tuan2013} using the same time-dependent
real-space Kubo method. In contrast to the long-range Thomas--Fermi
potential considered here, the onsite potential in Ref. \onlinecite{Tuan2013}
is set to zero and the scattering is due to grain boundaries separating
domains with different crystallographic orientations. Even though
this study did not discuss a functional dependence of the conductivity,
a visual inspection of the obtained results reveals an approximate
linear dependence of the conductivity on the Fermi energy, which is
consistent with our results. Moreover, the conductance of graphene
with several types of domain boundaries has also been shown to be
a linear function of the Fermi energy.\cite{Peres} We therefore speculate
that the linear energy dependence (and thus the sublinear density
dependence) of the conductivity is related to scattering off extended
defects. More systematic studies of scattering for different forms
of potentials are needed in order to clarify this question.

\section{Boltzmann approach\label{sec:Boltzmann_Approach}}

\subsection{Formalism}

In this section we tackle the problem of dc transport in graphene
with 1D charged defects by means of semiclassical Boltzmann theory.
We would like to stress that the full quantum calculations of Sec.~\ref{sec:TB_Kubo_RealSpace}
and semiclassical kinetic theory provide complementary information
about electronic transport; while the former is more suitable to handle
highly disordered systems or strong scattering regime (given practical
computational limitations),\cite{comment} semi-classical approaches
yield an accurate picture of charge transport for dilute disorder
and are often limited to the weak scattering regime (an exception
being resonant scattering which can be treated non-perturbatively).\cite{Ferreira}
Here, the dimensionless parameter $\beta\equiv|\Delta|L/(\hbar v_{F}$),
with $L$ of the order of the system size, defines the onset of weak
scattering regime, i.e.,~$\beta\ll1$. Note that the simulations
of the previous section have $\beta\gtrsim10^{2}$, and therefore
fall well inside the strong scattering regime.

The effective potential of a charged line is long-ranged and hence
we neglect intervalley scattering. Within the Dirac cone approximation,
the semiclassical dc conductivity of graphene at zero temperature
is given by\cite{Castro Netto review,Peres_review,DasSarma_review,Ferreira}
\begin{equation}
\sigma=\frac{ge^{2}}{2h}k_{F}v_{F}\tau(k_{F})\,.\label{eq:semiclassical_cond}
\end{equation}
In the above, the factor $g=4$ accounts for spin and valley degeneracies,
and $\tau(k_{F})$ is the transport scattering time at the Fermi surface
\begin{equation}
\tau(k_{F})=\left[n_{l}v_{F}\int d\theta(1-\cos\theta)|f(\theta)|^{2}\right]^{-1}\,,\label{eq:transport scattering time}
\end{equation}
where $f(\theta)$ is the scattering amplitude at an angle $\theta$
and $n_{l}$ stands for the (areal) density of charged lines.

In this work we compute the scattering amplitudes in the second Born
approximation (SBA) with respect to the scattering potential $V(\boldsymbol{r})$.
This allows us to improve over the commonly employed first Born approximation
(FBA) by capturing the non-trivial effect of electron--hole asymmetry;
see Fig.~\ref{Fig_FBA_vs_SBA}. In the Appendix C we show that the
SBA scattering amplitude for 2D massless fermions is given by 
\begin{eqnarray}
f_{\textrm{SBA}}(\theta) & = & \frac{\Xi(\theta)}{v_{F}\hbar}\sqrt{\frac{k}{8\pi}}\left\{ \tilde{V}(\mathbf{q})+\int\frac{d^{2}\mathbf{p}}{\left(2\pi\right)^{2}}\tilde{V}(\mathbf{k}^{\prime}-\mathbf{p})\times\right.\nonumber \\
 &  & \left.\langle u_{\mathbf{k}}|G_{0}(\mathbf{p})|u_{\mathbf{k}}\rangle\tilde{V}(\mathbf{p}-\mathbf{k})\right\} ,\label{eq:scattering_amplitude_SBA}
\end{eqnarray}
where $\tilde{V}(\mathbf{q})$ denotes the 2D Fourier transform of
the scattering potential energy of a charged line $\tilde{V}(\mathbf{q})=\int d^{2}\mathbf{r}e^{-i\mathbf{q}\cdot\mathbf{r}}V(\mathbf{r})$,
and $G_{0}(\mathbf{p})$ is the 2D Dirac fermion propagator for particles
with energy $E_{F}=sv_{F}\hbar k_{F},$ i.e., 
\begin{equation}
G_{0}(\mathbf{p})=\frac{1}{\hbar^{2}v_{F}^{2}}\frac{E_{F}+\hbar v_{F}\boldsymbol{\sigma}\cdot\mathbf{p}}{k_{F}^{2}-p^{2}+is0^{+}}\,.\label{eq:propagator}
\end{equation}
The symbol $s=\pm1$ distinguishes between electrons and holes, that
is, $s\equiv\textrm{sign}(E_{F})$. $\mathbf{k}$ is the wavevector
of the incident electron, $\hbar\mathbf{q}=\hbar(\mathbf{k}^{\prime}-\mathbf{k})$
is the transferred momentum ($\mathbf{k}^{\prime}$ stands for the
`out' wavevector), $\theta=\angle(\mathbf{k}^{\prime},\mathbf{k})$
is the scattering angle, $|u_{\mathbf{k}}\rangle=2^{-1/2}(1,\: se^{i\theta_{\mathbf{k}}})^{T}$
is the Dirac spinor for scattered particles, and the form factor $\Xi(\theta)=1+e^{i\theta}$
comes from graphene's sublattice symmetry and precludes carriers from
back-scatter. Without loss of generality, in what follows, we consider
incident carriers propagating along the $x$-direction, $\mathbf{k}=k_{F}e_{\boldsymbol{x}}$.

\begin{figure}
\centering{}\includegraphics[clip,width=1\columnwidth]{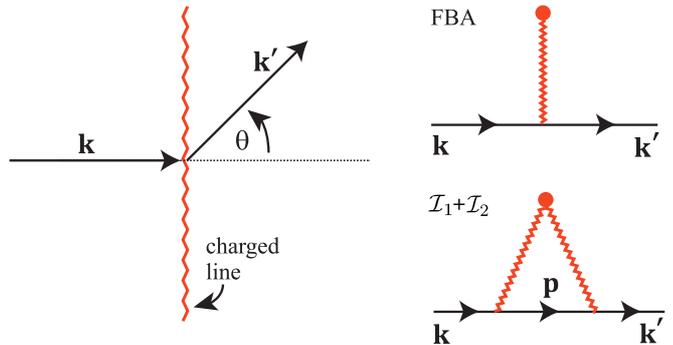}
\caption{\label{Fig_FBA_vs_SBA} (Color online) Schematic picture showing a
scattering event and Feynman scattering diagrams considered in this
work. The circle signifies the extended charged defect with charge
density $\lambda$ and the zigzag (solid) line denotes the scattering
potential (bare propagator). The transport relaxation rate is $\mathcal{O}(\lambda^{2})$
in the FBA approximation (top diagram) and $\mathcal{O}(\lambda^{3})$
in the SBA (bottom diagram). }
\end{figure}

The first term inside brackets in Eq.~(\ref{eq:scattering_amplitude_SBA})
is proportional to the Fourier transform of the scattering potential
evaluated at the transferred momentum $\hbar\mathbf{q}$, that is,
the familiar FBA scattering amplitude. The remaining terms result
from the next-order correction to the FBA and require the calculation
of two integrals, namely 
\begin{eqnarray}
 & \mathcal{I}_{1}\equiv\hbar v_{F}k_{F}\int\frac{d^{2}\mathbf{p}}{\left(2\pi\right)^{2}}\tilde{V}(\mathbf{k}^{\prime}-\mathbf{p})g(\mathbf{p})\tilde{V}(\mathbf{p}-\mathbf{k}),\label{eq:a1_SBA}\\
 & \mathcal{I}_{2}\equiv\int\frac{d^{2}\mathbf{p}}{\left(2\pi\right)^{2}}\tilde{V}(\mathbf{k}^{\prime}-\mathbf{p})[\hbar v_{F}\mathbf{p}\cdot\boldsymbol{e}_{x}]g(\mathbf{p})\tilde{V}(\mathbf{p}-\mathbf{k}).\label{eq:a2_SBA}
\end{eqnarray}
In writing these equations, we have defined the function 
\begin{equation}
g(\mathbf{p})=(\hbar v_{F})^{-2}\left(k_{F}^{2}-p^{2}+is0^{+}\right)^{-1}\,.\label{eq:gd}
\end{equation}

The scattering potential of an infinite line with density charge $\rho=\lambda\delta(x)\delta(z)$
was derived by some of the authors in Ref.~\onlinecite{FerreiraEPL}
and is given by 
\begin{equation}
\tilde{V}(\boldsymbol{q})=2\pi\delta(q_{y})\frac{\Delta}{|q_{x}|+q_{\textrm{TF}}}\,,\label{eq:potential_energy_infinite_line}
\end{equation}
where the parameter with units of energy $\Delta$ relates to the
charge density of a line $\lambda$ according to $\Delta=s\lambda e/(2\varepsilon_{0})$
(in vacuum); note that the absolute value of $\Delta$ coincides with
the definition of $\Delta$ as given in Sec.~\ref{sub:sub_Basics}.
The delta function in Eq.~(\ref{eq:potential_energy_infinite_line})
reflects momentum conservation along the direction defined by the
line. For completeness, a derivation of this result is provided in
Appendix~A.

In order to mimick the effect of lines with finite length we have
to modify Eq.~(\ref{eq:potential_energy_infinite_line}) as to allow
for momentum transfer to occur along both spatial directions. To this
end, we introduce a length scale associated with the line's average
length $L$. In the limit of small $k_{F}L$, we replace $2\pi\delta(q_{y})\rightarrow L$,\cite{finite_line}
as to obtain 
\begin{equation}
\tilde{V}_{L}(\boldsymbol{q})\equiv\frac{L\Delta}{|q_{x}|+q_{\textrm{TF}}}\,.\label{eq:potential_finite_line}
\end{equation}
We use this potential as a toy model for describing transport for
dilute concentrations of lines of charge. The particularly simple
form of $\tilde{V}_{L}(\boldsymbol{q})$ allows for an exact calculation
of scattering amplitudes, as shown in what follows.

\subsection{First Born approximation}

\begin{figure}
\centering{}\includegraphics[clip,width=1\columnwidth]{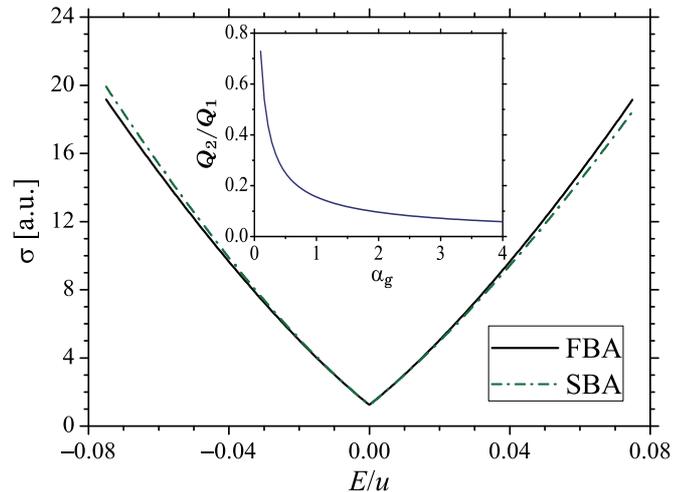}
\caption{\label{Fig_FBA_vs_SBA_2} (Color online) Semi-classical dc conductivity
at fixed Thomas--Fermi wavevector as a function of Fermi energy in
the weak scattering regime with $\beta=0.08$ and $q_{\textrm{TF}}a=0.01$.
Inset: The ratio $Q_{2}(\alpha_{\textrm{g}})/Q_{1}(\alpha_{\textrm{g}})$
determining the amount of transport electron--hole asymmetry is plotted
as function of graphene's effective structure factor $\alpha_{\textrm{g}}$.}
\end{figure}

The FBA provides a good approximation to transport scattering rates
for 1D charged defects with $|\Delta|\ll\hbar v_{F}L^{-1}$ ($\beta\ll1$).
Within the FBA we retain only the first term in Eq.\,(\ref{eq:scattering_amplitude_SBA}).
The transport relaxation rate 
\begin{eqnarray}
[\tau_{\textrm{FBA}}(k_{F})]^{-1} & = & n_{l}v_{F}\frac{L^{2}\Delta^{2}k_{F}}{8\pi v_{F}^{2}\hbar}\int_{0}^{2\pi}d\theta(1-\cos\theta)\times\nonumber \\
 &  & |\Xi(\theta)|^{2}[k_{F}(1-\cos\theta)+q_{\textrm{TF}}]^{-2}\,,\label{eq:rate_FBA}
\end{eqnarray}
can be computed analytically and leads to the following result 
\begin{equation}
\tau_{\textrm{FBA}}(k_{F})=\frac{1}{v_{F}n_{l}\beta^{2}}\frac{2k_{F}\sqrt{q_{\textrm{TF}}(2k_{F}+q_{\textrm{TF}})}}{k_{F}+q_{\textrm{TF}}-\sqrt{q_{\textrm{TF}}(2k_{F}+q_{\textrm{TF}})}}\,.\label{eq:tau_FBA}
\end{equation}
Invoking the semi-classical expression for the dc conductivity Eq.~(\ref{eq:semiclassical_cond})
and the relation $q_{\textrm{TF}}=4\alpha_{\textrm{g}}k_{F}$, we
conclude that $\sigma$ is proportional to $k_{F}^{2}$, according
to 
\begin{equation}
\sigma=\frac{4e^{2}}{h}\frac{f(4\alpha_{\textrm{g}})}{n_{l}\beta^{2}}k_{F}^{2}\,,\label{eq:sigma_FBA}
\end{equation}
with $f(x)=g(x)/[1+x-g(x)]$ and $g(x)=\sqrt{x(2+x)}$. The dependence
of Eqs.~(\ref{eq:tau_FBA})--(\ref{eq:sigma_FBA}) on the Fermi wavevector
could be anticipated from the form of the effective potential in Fourier
space, Eq.~(\ref{eq:potential_finite_line}); as $\tilde{V}\propto k_{F}^{-1}$,
the relaxation time must be proportional to $k_{F}$ at all orders
in the Born series, implying $\sigma\propto k_{F}^{2}\propto|n_{e}|$.
In other words, higher order corrections to the FBA renormalize the
mobility of carriers $\mu\equiv\sigma/|n_{e}|$, while preserving
the overall dependence of $\sigma$ and $\mu$ on the Fermi energy.
This property is specific to the potential $\tilde{V}_{L}(\boldsymbol{q})$
and therefore is not expected to hold in models of extended charged
defects beyond the limit of small $k_{F}L$.

We briefly discuss how the FBA conductivity compares with the results
reported earlier in Fig.~\ref{Fig_Cond_posit-negat_posit}. The solid
line in Fig.~\ref{Fig_FBA_vs_SBA_2} shows the FBA conductivity at
fixed Thomas--Fermi wavevector, i.e.,~$q_{\textrm{TF}}a=0.01$. In
this case, the function $f(4\alpha_{\textrm{g}})$ is no longer constant
and the functional dependence of $\sigma$ with $k_{F}$ is changed
to linear at small energies, hence resembling the Kubo results. However,
this comparison should not be pushed too far; note that the slope
of the FBA $\sigma$ versus $E$ curve depends linearly on $q_{\textrm{TF}}$,
and therefore Eq.~(\ref{eq:sigma_FBA}) (and hence the quadratic
dependence) is recovered when the self-consistent relation $q_{\textrm{TF}}=4\alpha_{\textrm{g}}k_{F}$
is used. In fact, 
\begin{equation}
\sigma(k_{F},q_{\textrm{TF}})=\frac{4e^{2}}{h}\frac{q_{\textrm{TF}}k_{F}}{n_{l}\beta^{2}},\quad k_{F}\ll q_{\textrm{TF}}\,.\label{eq:sigma_FBA_fixed_QTF}
\end{equation}
whereas the Kubo simulations show $\sigma\propto k_{F}$ independently
on $q_{\textrm{TF}}$ in a wide range of energies. The latter behavior
does not occur in the weak scattering regime described here.

\subsection{Second Born approximation: electron--hole asymmetry}

A large sensitivity to the carriers polarity in transport dominated
by charged lines is borne out in the numerical simulations of Sec.~\ref{sec:TB_Kubo_RealSpace}.
Here, we describe this effect from the point of view of semiclassical
transport theory. According to the Fermi golden rule the transport
relaxation rate depends on the modulus square of the scattering potential;
hence, in FBA approximation, opposite charges $\pm e$ have the same
scattering amplitudes and hence cannot be distinguished. As observed
earlier, the dependence of $\sigma$ on the carriers polarity can
be captured by retaining the next term in the Born series for the
scattering amplitude $f(\theta)$---the SBA bottom diagram in Fig.~\ref{Fig_FBA_vs_SBA}.

We compute the transport electron--hole asymmetry, defined as 
\begin{equation}
\delta\equiv\left|\frac{\sigma_{\textrm{dc}}-\sigma_{\textrm{dc}}^{*}}{\sigma_{\textrm{dc}}+\sigma_{\textrm{dc}}^{*}}\right|\,,\label{eq:e_h_assymetry}
\end{equation}
with $\sigma_{\textrm{dc}}^{*}\equiv\left.\sigma_{\textrm{dc}}\right|_{e\rightarrow-e}=\left.\sigma_{\textrm{dc}}\right|_{\Delta\rightarrow-\Delta}$.
In the weak scattering regime, $L|\Delta|\ll\hbar v_{F}$, the asymmetry
parameter is proportional to the ratio of the bottom to the top diagrams
in Fig.~\ref{Fig_FBA_vs_SBA}. Explicitly, 
\begin{equation}
\delta=\frac{2\,\textrm{sign}\Delta\,\textrm{Re}\int d\theta|\Xi(\theta)|^{2}(1-\cos\theta)[\mathcal{I}_{1}+\mathcal{I}_{2}]\tilde{V}(q_{\theta})}{\int d\theta|\Xi(\theta)|^{2}(1-\cos\theta)|\tilde{V}(q_{\theta})|^{2}}\,,\label{eq:assym_param}
\end{equation}
where $q_{\theta}=2k_{F}\sin(\theta/2)$ is the transferred momentum
in elastic scattering events. Remark that $\mathcal{I}_{1(2)}$ in
Eq.~(\ref{eq:assym_param}) depend on the angle $\theta$ through
the wavevector $\mathbf{k}^{\prime}$ {[}c.f.,~Eqs.~(\ref{eq:a1_SBA})--(\ref{eq:a2_SBA}){]}.
The derivation of this and related results is given in Appendix~D.

Inserting the potential energy of a charged line Eq.~(\ref{eq:potential_finite_line})
into the above expression and performing the angular integration yields
\begin{equation}
\delta\approx\left|\frac{\Delta L}{\hbar v_{F}}\right|\frac{Q_{2}(\alpha_{\textrm{g}})}{Q_{1}(\alpha_{\textrm{g}})}\,.\label{eq:delta}
\end{equation}
The explicit form of the functions $Q_{1}(\alpha_{\textrm{g}})$ and
$Q_{2}(\alpha_{\textrm{g}})$ is given in Eqs.~(\ref{eq:I1_exp})
and (\ref{eq:I2}), respectively. For the toy model of a charged line
considered here {[}Eq.~(\ref{eq:potential_finite_line}){]}, cross
sections are proportional to $k_{F}^{-1}$ at all orders, implying
that the asymmetry parameter $\delta$ is insensitive to the Fermi
energy. Indeed, the electron--hole asymmetry depends only on the magnitude
of the Thomas--Fermi screening through the effective graphene's structure
constant, $\alpha_{\textrm{g}}$. In vacuum, $\alpha_{\textrm{g}}\approx2.5$,
and the evaluation of Eq.~(\ref{eq:delta}) yields $\delta\approx0.08\cdot\beta$.
The ratio $Q_{2}/Q_{1}$ is found to be very sensitive to the effective
screening length of a charged line (refer to inset of Fig.~\ref{Fig_FBA_vs_SBA_2});
for $q_{\textrm{TF}}\gg k_{F}$ ($\alpha_{\textrm{g}}\gg1$) screening
is very efficient and electron--hole asymmetry is negligible, whereas
for $q_{\textrm{TF}}\lesssim k_{F}$ ($\alpha_{\textrm{g}}\lesssim1$)
the ratio $Q_{2}/Q_{1}$ can assume large values leading to an enhancement
of the asymmetry parameter $\delta$.

The transport electron--hole asymmetry in scattering events reflects
into a decrease (increase) of the SBA transport relaxation time with
respect to the FBA result for positive (negative) Fermi energy. In
fact, by expanding the SBA transport relaxation rate Eq.~(\ref{eq:transport scattering time})
in the small parameter $\beta$, we find 
\begin{equation}
\tau(k_{F})=\left[1-s\delta(\alpha_{\textrm{g}})+\mathcal{O}(\beta^{2})\right]\tau_{\textrm{FBA}}(k_{F}).\label{eq:transport_scatt_time_assym}
\end{equation}
This result shows that the effect of second term in the Born series
(bottom diagram in Fig.~\ref{Fig_FBA_vs_SBA}) is to renormalize
the transport relaxation time according to the carriers polarity,
$s$, and screening strength $\alpha_{\textrm{g}}$. This behavior
is qualitatively consistent with the numerical Kubo simulations (see
Fig.~\ref{Fig_Cond_posit-negat_posit}, for instance). In order to
make the comparison between the semiclassical SBA prediction and the
simulations shown in Sec.~\ref{sec:TB_Kubo_RealSpace} more accurate,
we investigate the behavior of Eq.~(\ref{eq:transport_scatt_time_assym})
at fixed Thomas--Fermi wavevector. Note that, in this case, the asymmetry
parameter becomes a function of the Fermi energy according to $\delta=\delta(q_{\textrm{TF}}/4k_{F})$.
Given the behavior of the function $Q_{2}/Q_{1}$ at small values
of its argument (see inset of Fig.~\ref{Fig_FBA_vs_SBA_2}), the
asymmetry at fixed $q_{\textrm{TF}}$ can be quite large even at modest
$k_{F}$, originating a considerable deviation of the conductivity
at fixed $q_{\textrm{TF}}$ from its FBA value, as depicted in the
main panel of Fig.~\ref{Fig_FBA_vs_SBA_2}.

\subsection{Comparison with Kubo simulations\label{sec:Comparison}}

\emph{Variation of conductivity with electronic density}. In the strong
scattering regime, the numerical Kubo simulations disclose a dc conductivity
that is linear in the Fermi wavevector, $\sigma\propto k_{F}\propto|n_{e}|^{1/2}$,
a very distinct behavior from the semiclassical prediction for the
dc conductivity, $\sigma\propto k_{F}^{2}\propto|n_{e}|$. At first
sight, it seems that both results are irreconcilable; after all they
focus on opposite scattering regimes. However, for the toy model of
a charged line considered here {[}Eq.~(\ref{eq:potential_finite_line}){]},
$\sigma\propto k_{F}^{2}$ at all orders in perturbation theory, and
hence we would expect similar semiclassical behavior even in the strong
scattering regime. In order to investigate this question further,
we have performed numerical Kubo simulations for a dilute system with
a single line of charge in the strong scattering regime (see Appendix
B). These simulations show the same functional dependence $\sigma=\sigma(n_{e})$
than the simulations of Sec.~\ref{sec:TB_Kubo_RealSpace} for highly
disordered configurations. This indicates a possible failure of the
toy model in describing the potential landscape of the simulations
in a wider range of electronic densities; remark that, by construction,
Eq.~(\ref{eq:potential_finite_line}) should provide a good description
of transport only at low Fermi momentum.

\emph{Transport electron--hole asymmetry}. A decrease (increase) of
the electronic mobility for electrons (holes) with respect to the
particle--hole symmetric case $V\gtrless0$ is found in all numerical
simulations with $V>0$ (Figs.~\ref{Fig_Cond_posit-negat_posit}
and \ref{Fig_self-cons_cond}). This effect can be ascribed to the
shift of the charge neutrality point towards positive energy values
caused by a potential landscape with positive sign (see density of
states in Fig.~\ref{Fig_DOS}). Although the semiclassical picture
is build upon the density of states of bare graphene, the inclusion
of higher-order diagrams (Fig.~\ref{Fig_FBA_vs_SBA}) in the calculation
of the scattering amplitude renormalizes the relaxation rates according
to the carriers polarity, thus accounting correctly for the general
behavior of the transport electron--hole asymmetry.

\section{Conclusions\label{sec:Conclusions}}

In this work we have considered theoretically the transport properties
of graphene with extended charged defects. Recent experiments show
that these defects are ubiquitous in chemically synthesized graphene
systems and degrade their electronic mobilities. We modeled extended
charged defects by lines with uniform charge densities and computed
their potentials according to a self-consistent Thomas--Fermi approach.
In contrast to the charged point defects, the potential of a line
of charge is screened poorly by low-energy excitations in graphene,
resulting in long-ranged effective potentials. We considered the regimes
of weak and strong scattering by means of semiclassical Boltzmann
theory and large-scale numerical evaluation of the Kubo formula, respectively.
Whereas the semiclassical calculation reveals a familiar linear dependence
of conductivity with the electronic density, the Kubo simulations
show a robust sublinear dependence and conductivity nearly constant
by varying the Thomas--Fermi wavelength by almost one order of magnitude.
The latter is a remarkable property of extended charged defects in
graphene. 
\begin{acknowledgments}
T.M.R., A.A.S., and I.V.Z. gratefully acknowledge financial support
from the Swedish Institute and thank Stephan Roche for discussions
about the time-dependent Kubo approach. A.F. greatly acknowledges
support from National Research Foundation\textendash{}Competitive
Research Programme through award `Novel 2D materials with tailored
properties: beyond graphene' (Grant No. R-144-000-295-281) and discussions
with N.M.R. Peres and M.A. Cazalilla. 
\end{acknowledgments}
\appendix

\section*{Appendix A: Thomas--Fermi renormalized potential of a charged line
in graphene and respective fitting by a Lorentzian function}

Here we derive the effective potential of an infinite charged line
within the Thomas--Fermi (TF) approximation. Rearrangements of electronic
density in a metal, around an impurity, does not alter the Fermi energy
$E_{F}$, and thus we may write\cite{Ziman} 
\begin{equation}
E_{F}\simeq\epsilon(\mathbf{r})-e\varphi_{\textrm{2D}}(\mathbf{r})\,,\label{eq:1}
\end{equation}
where $\epsilon(\mathbf{r})$ and $e\varphi_{\textrm{2D}}(\mathbf{r})$
are, respectively, the local energy of the electrons at the top of
the band and the effective potential energy induced by the impurity
charge. In our problem the metal is graphene (at finite densities)
and the impurity is a charged line. Let $n_{\text{eq}}$ be the electronic
density of pristine graphene, then 
\begin{equation}
\epsilon(\mathbf{r})=E_{F}+\left.\frac{d\epsilon}{dn(\mathbf{r})}\right|_{n(\mathbf{r})-n_{\text{eq}}}\left[n(\mathbf{r})-n_{\text{eq}}\right]\label{eq:2}
\end{equation}
to first order in $\delta n(\mathbf{r})\equiv n(\mathbf{r})-n_{\text{eq}}$.
We thus arrive at the following relation between the potential energy
and the charge density 
\begin{eqnarray}
e\varphi_{\textrm{2D}}(\mathbf{r}) & \simeq & \left.\frac{d\epsilon}{dn(\mathbf{r})}\right|_{n(\mathbf{r})=n_{\text{eq}}}\left[n(\mathbf{r})-n_{\text{eq}}\right]\notag\\
 & = & \frac{\kappa}{2\sqrt{n(\mathbf{r})}}\delta n(\mathbf{r}),\label{eq:3}
\end{eqnarray}
where $\epsilon(r)\simeq\kappa\sqrt{n(\mathbf{r})}$ with $\kappa=\hbar v_{F}k_{F}/\sqrt{n_{\text{eq}}}$.

The above equations show that in order to maintain the Fermi level
constant, a change in the local electronic density takes place. The
effective potential has to be determined self-consistently solving
the Poisson's equation. According to the TF approximation, we have
\begin{equation}
\nabla^{2}\varphi_{\textrm{eff}}(\mathbf{r},z)=-\frac{1}{\varepsilon_{0}\varepsilon_{\textrm{r}}}\left[\rho_{\text{imp}}(\mathbf{r},z)+\delta\rho(\mathbf{r},z)\right],\label{eq:5}
\end{equation}
where $\varepsilon_{0}$ $(\varepsilon_{\textrm{r}})$ is a vacuum
(relative) permittivity. Note that in the above equation $\varphi_{\textrm{eff}}(\mathbf{r},z)$
depends on the in-plane coordinates $\mathbf{r}$ and $z$. We consider
a line defect with charge per unit of length $\lambda$, and orientated
along the $y$-axis, 
\begin{equation}
\rho_{\text{imp}}(\mathbf{r},z)=\lambda\delta(x)\delta(z).\label{eq:6}
\end{equation}
From Eqs.~(\ref{eq:5})--(\ref{eq:6}) and 
\begin{align}
\delta\rho(\mathbf{r},z) & =-e\delta n(\mathbf{r})\delta(z)\,\nonumber \\
 & =-\frac{2e^{2}}{\kappa}\sqrt{n_{\text{eq}}}\varphi_{\textrm{2D}}(\mathbf{r})\delta(z)\,,\label{eq:delta_rho}
\end{align}
we arrive at the important intermediate result 
\begin{align}
\nabla^{2}\varphi_{\textrm{eff}}(\mathbf{r},z)= & \frac{1}{\varepsilon_{0}\varepsilon_{\textrm{r}}}\left[\frac{2e^{2}}{\kappa}\sqrt{n_{\text{eq}}}\varphi_{\textrm{2D}}(\mathbf{r})-\lambda\delta(x)\right]\delta(z).\label{eq:7}
\end{align}
Note that the term $\delta\rho(\mathbf{r})$ in Eq.~(\ref{eq:5})
is not only a self-consistent term, but also imposes an important
geometric restriction by forcing the rearrangement of charge to occur
in the graphene plane. We solve the Poisson equation Eq.~(\ref{eq:8})
using the Fourier transform method, viz., 
\begin{equation}
\left(q_{x}^{2}+q_{z}^{2}\right)\varphi_{\textrm{eff}}(q_{x},q_{z})=\frac{\lambda}{\varepsilon_{0}\varepsilon_{\textrm{r}}}-2q_{\textrm{TF}}\varphi_{\textrm{2D}}(q_{x})\,,\label{eq:8}
\end{equation}
where we have defined $q_{\textrm{TF}}=e^{2}\sqrt{n_{\text{eq}}}/(\varepsilon_{0}\varepsilon_{\textrm{r}}\kappa)=4\alpha_{\textrm{g}}k_{F}$.
Integrating out the $q_{z}$ dependence leads to 
\begin{eqnarray}
\varphi_{\textrm{2D}}(q_{x}) & \equiv & \int\frac{dq_{z}}{2\pi}\varphi_{\text{eff}}(q_{x},q_{z})=\frac{\lambda/(2\varepsilon_{0}\varepsilon_{\textrm{r}})}{q_{\textrm{TF}}+|q_{x}|}.\label{eq:9}
\end{eqnarray}
The effective potential in a real space is therefore given by 
\begin{equation}
\varphi(x)\equiv\varphi_{\textrm{2D}}(x)=\frac{\lambda}{2\varepsilon_{0}\varepsilon_{\textrm{r}}}\int_{0}^{\infty}\frac{dq_{x}}{\pi}\frac{\cos\left(q_{x}x\right)}{q_{\textrm{TF}}+q_{x}}\,,\label{eq:10}
\end{equation}
or, equivalently, 
\begin{align}
\varphi(x)= & \frac{\lambda}{2\pi\varepsilon_{0}\varepsilon_{\textrm{r}}}\Bigl\{-\cos\left(q_{\textrm{TF}}x\right)\text{Ci}\left(q_{\textrm{TF}}x\right)+\notag\\
 & +\sin\left(q_{\textrm{TF}}x\right)\left[\frac{\pi}{2}-\text{Si}\left(q_{\textrm{TF}}x\right)\right]\Bigr\},\label{eq:effective_potential_real_space}
\end{align}
where $\text{Ci}$ and Si denote the cosine and sine integral functions.
The above equation possesses the following asymptotic behavior: 
\begin{equation}
\varphi(x)\longrightarrow\begin{cases}
\frac{\lambda}{2\pi\varepsilon_{0}\varepsilon_{\textrm{r}}}\left(\frac{1}{q_{\textrm{TF}}x}\right)^{2},\, & q_{\textrm{TF}}x\gg1,\\
\frac{\lambda}{2\pi\varepsilon_{0}\varepsilon_{\textrm{r}}}\ln\left(\frac{1}{q_{\textrm{TF}}x}\right),\, & q_{\textrm{TF}}x\ll1.
\end{cases}\label{eq:effective_potential_real_space_asympt}
\end{equation}

\begin{figure}
\centering{}\includegraphics[width=0.7\columnwidth]{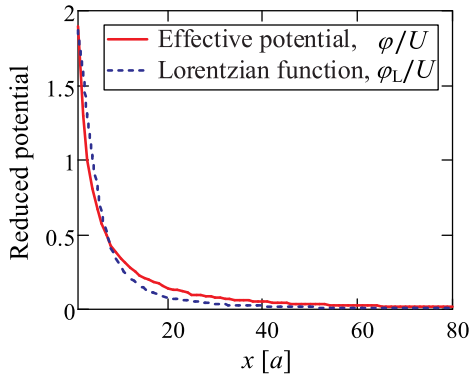} \caption{\label{Fig_Lorentzian} (Color online) The Thomas--Fermi potential
(\ref{eq:effective_potential_real_space}) fitted by the Lorentzian
function (\ref{Lorentzian}). Here, $q_{\textrm{TF}}a=0.1$, $U=\lambda/(2\pi\varepsilon_{0}\varepsilon_{\textrm{r}})$,
and the fitting parameters are $A=1.544$, $B=0.780$, $C=0.046$.}
\end{figure}

The obtained expression for the effective potential, Eq.~(\ref{eq:effective_potential_real_space}),
is well fitted by the Lorentzian function, 
\begin{equation}
\varphi_{\textrm{L}}(x)=\frac{\lambda}{2\pi\varepsilon_{0}\varepsilon_{\textrm{r}}}\frac{A}{B+Cx^{2}},\label{Lorentzian}
\end{equation}
where fitting parameters $A$, $B$, $C$ can be calculated from the
least-squares method, see Fig.~\ref{Fig_Lorentzian}. We use Eq.~(\ref{Lorentzian})
in the numerical calculation based on the Kubo approach.

\section*{Appendix B: Self-consistent calculations of the conductivity for
a single charged line}

\begin{figure*}
\centering{}\includegraphics[width=1\textwidth]{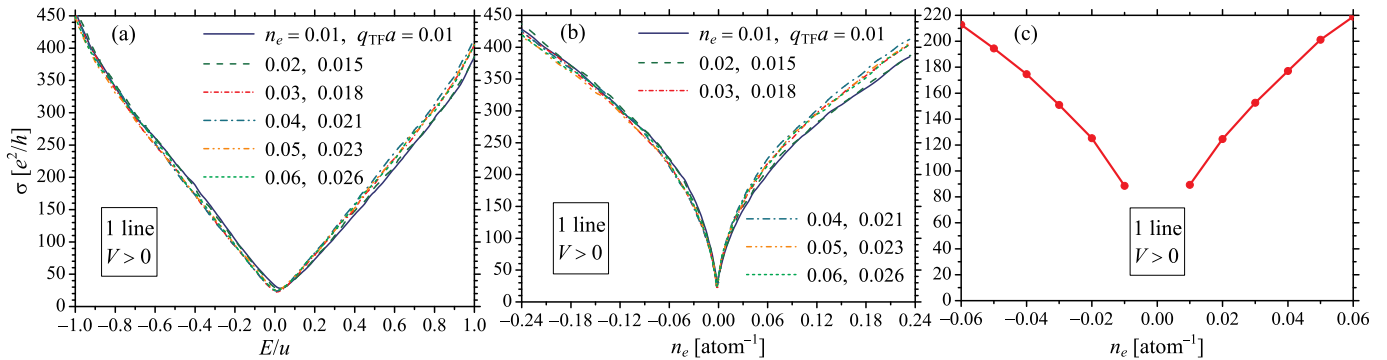} \caption{\label{Fig_self-cons_cond} (Color online) Conductivity vs. the energy
(a) and electron density (b), (c) for nonself-consistent (a), (b)
and self-consistent (c) effective potentials $V\sim U=\lambda/(2\pi\varepsilon_{0}\varepsilon_{\textrm{r}})\in[0,\triangle]$
($\triangle=0.25u$) describing a single charged line. Curve in (c)
is plotted combining corresponding values for $\sigma$ on curves
in (b). The correspondence between $q_{\textrm{TF}}$ used in calculations
of the potential and respective $n_{e}$ is indicated in (a) and (b).}
\end{figure*}

The Thomas--Fermi wavevector $q_{\textrm{TF}}$ entering the effective
scattering potential {[}Eq.~(\ref{Eq_TF potential}) or (\ref{eq:effective_potential_real_space}){]}
depends on the electron density $n_{e}$. In this appendix we check
how this dependence affects the behavior of $\sigma$ as compared
with the results obtained in Sec.~\ref{sec:TB_Kubo_RealSpace} for
a fixed $q_{\textrm{TF}}$; accounting for the density dependence
makes our effective potential `self-consistent'.

We perform our calculations as follows. In the Kubo method used in
this study it is not possible to change the scattering potential while
changing the energy (or density) of the electrons. We therefore perform
independent calculations for six different values of $q_{\textrm{TF}}$
obtaining six different dependencies $\sigma=\sigma(E)$ and $\sigma=\sigma(n_{e})$
as shown in Fig.~\ref{Fig_self-cons_cond}(a) and Fig.~\ref{Fig_self-cons_cond}(b),
respectively. In each dependence $\sigma=\sigma(n_{e})$ we choose
only one particular point (for both \textit{n}- and \textit{p}-types
of charge carriers) where the electron density $n_{e}$ corresponds
to $q_{\textrm{TF}}$ used in the calculation of this dependence (recall
that $q_{\textrm{TF}}$ scales as $q_{\textrm{TF}}\propto\sqrt{|n_{e}|}$).
Combining these six points on a single plot yields a `self-consistent'
curve $\sigma=\sigma(n_{e})$ as shown in Fig.~\ref{Fig_self-cons_cond}(c).
Figure~\ref{Fig_self-cons_cond} clear demonstrates that energy and
electron density dependencies of conductivity exhibit respectively
linear and sublinear behaviors, which are the same as corresponding
behaviors of the conductivities for the case of a fixed $q_{\textrm{TF}}$,
see Fig.~\ref{Fig_Cond_posit-negat_posit}. Note that electron--hole
asymmetry in Fig.~\ref{Fig_self-cons_cond} is weak since the source
of disorder here is due to a single line only (c.f.,~with 10 and
50 lines in Fig.~\ref{Fig_Cond_posit-negat_posit}).

\section*{Appendix C: Scattering amplitudes in the second Born approximation}

The scattering problem $(\hat{H}_{0}+\hat{V}-E)\Psi_{\mathbf{k}}=0$,
where $H_{0}$ denotes the free Hamiltonian and $\hat{V}$ a potential,
has the formal solution 
\begin{equation}
\Psi_{\mathbf{k}}=\phi_{\mathbf{k}}+\hat{G}_{0}\hat{V}\Psi_{\mathbf{k}}\,,\label{eq:Lippmann_Schwinger}
\end{equation}
where $\phi_{\mathbf{k}}$ solves the free Schrödinger equation $(\hat{H}_{0}-E)\phi_{\mathbf{k}}=0$
and describes the state of the incident particles. The resolvent is
given by $\hat{G}_{0}(z)=1/(z-\hat{H}_{0})$, where $z$ includes
an infinitesimally small imaginary part.

In the context of the present work, $H_{0}$ stands for the Hamiltonian
of pristine graphene in the single Dirac cone approximation, and $\hat{V}$
refers to the potential of a charged 1D defect (Appendix A). Although
the form of $\hat{V}$ remains unspecified in what follows it is assumed
to be a scalar in both sublattice and spin spaces. The spinor $\phi_{\mathbf{k}}(\mathbf{r})\equiv\langle\mathbf{r}|\phi_{\mathbf{k}}\rangle$
has the form\cite{Castro Netto review,Peres_review,DasSarma_review}
\begin{equation}
\phi_{\mathbf{k}}(\mathbf{r})=u_{\mathbf{k}}e^{i\mathbf{k}\cdot\mathbf{r}}\,,\label{eq:solution_slg}
\end{equation}
with 
\begin{equation}
u_{\mathbf{k}}=\frac{1}{\sqrt{2}}\left(\begin{array}{c}
1\\
se^{i\theta_{\mathbf{k}}}
\end{array}\right)\,.\label{eq:spinor_slg}
\end{equation}
In the above, $\theta_{\mathbf{k}}\equiv\arctan(k_{y}/k_{x})$ and
$s\equiv\textrm{sign}\left(E\right)$. Switching Eq.~(\ref{eq:Lippmann_Schwinger})
to the position representation, we obtain the Lippmann--Schwinger
equation 
\begin{equation}
\Psi_{\mathbf{k}}(\mathbf{r})=\phi_{\mathbf{k}}(\mathbf{r})+\int d^{2}\mathbf{r}^{\prime}G_{0}(\mathbf{r}-\mathbf{r}^{\prime})V(\mathbf{r}^{\prime})\Psi_{\mathbf{k}}(\mathbf{r^{\prime}})\,,\label{eq:LS_position_rep}
\end{equation}
where $G_{0}(\mathbf{r}-\mathbf{r}^{\prime})=\langle\mathbf{r}|\hat{G}_{0}(z)|\mathbf{r}^{\prime}\rangle$
is the Green function of the problem. The graphene Hamiltonian reads
\begin{equation}
\hat{H}_{0}=\hbar v_{F}\boldsymbol{\sigma}\cdot\hat{\mathbf{p}}\,,\label{eq:Dirac}
\end{equation}
and the Fourier transform of the Green function $G_{0}(\mathbf{p})=\int d^{2}\mathbf{r}\exp\left(-i\mathbf{p}\cdot\mathbf{r}\right)G_{0}(\mathbf{r})$
is given by 
\begin{equation}
G_{0}(\mathbf{p})=\left(z-\hbar v_{F}\boldsymbol{\sigma}\cdot\mathbf{p}\right)^{-1}\,.\label{eq:green_function}
\end{equation}
In what follows, unless stated otherwise, we set $\hbar\equiv1\equiv v_{F}$
. It is also convenient to recast Eq.~(\ref{eq:green_function})
in the form 
\begin{eqnarray}
G_{0}(\mathbf{p}) & = & g(\mathbf{p})(z+\boldsymbol{\sigma}\cdot\mathbf{p})\,,\label{eq:green_function_aux1}\\
g(\mathbf{p}) & = & \left(z^{2}-p^{2}\right)^{-1}.\label{eq:green_function_aux2}
\end{eqnarray}
where $z=E+is0^{+}$; the inclusion of a small imaginary part $is0^{+}$
amounts to consider outgoing waves (see below). For simplicity we
focus on scattering of positive energy carriers (electrons), $s=1$.
We write $E=k$ and evaluate the Green function in real space representation
\begin{align}
G_{0}(\mathbf{r}-\mathbf{r}^{\prime}) & =\left(E-i\boldsymbol{\sigma}\cdot\mathbf{\nabla}\right)\int\frac{d^{2}\mathbf{p}}{\left(2\pi\right)^{2}}e^{i\mathbf{p}\cdot(\mathbf{r}-\mathbf{r}^{\prime})}g(\mathbf{p})\label{eq:GF_1}\\
 & =-\frac{i}{4}\left(k-i\boldsymbol{\sigma}\cdot\mathbf{\nabla}\right)H_{0}^{(1)}\left(k|\mathbf{r}-\mathbf{r}^{\prime}|\right)\,,\label{eq:GF_2}
\end{align}
where $H_{n}^{(1)}\left(k|\mathbf{r}-\mathbf{r}^{\prime}|\right)$
is the first kind Hankel function of order $n$, whose asymptotic
form is that of outgoing cylindrical waves. Using the property $\partial_{x}H_{0}^{(1)}(x)+H_{1}^{(1)}(x)=0$,
the second term in Eq.~(\ref{eq:GF_2}) can be written in the simple
form 
\begin{equation}
\boldsymbol{\sigma}\cdot\mathbf{\nabla}H_{0}^{(1)}(k|\mathbf{r}-\mathbf{r}^{\prime}|)=-kH_{1}^{(1)}(k|\mathbf{r}-\mathbf{r}^{\prime}|)\sigma_{\theta}\,,\label{eq:identity_Hankel}
\end{equation}
where we have introduced the matrix 
\begin{equation}
\sigma_{\theta}\equiv\left(\begin{array}{cc}
0 & e^{-i\theta}\\
e^{i\theta} & 0
\end{array}\right)\,.\label{eq:sigma_theta}
\end{equation}
In the above, the angle $\theta\equiv\theta(\mathbf{r},\mathbf{r}^{\prime})$
is defined through the relation $\left(\mathbf{r}-\mathbf{r}^{\prime}\right)/|\mathbf{r}-\mathbf{r}^{\prime}|=\left(\cos\theta,\sin\theta\right)^{T}$.

Combining Eqs.~(\ref{eq:GF_2})--(\ref{eq:identity_Hankel}) we obtain
the explicit form of the Green function of pristine graphene 
\begin{equation}
G_{0}(\mathbf{r}-\mathbf{r}^{\prime})=-\frac{ik}{4}\left[H_{0}^{(1)}(k|\mathbf{r}-\mathbf{r}^{\prime}|)+i\sigma_{\theta}H_{1}^{(1)}(k|\mathbf{r}-\mathbf{r}^{\prime}|)\right]\,.\label{eq:GF_position}
\end{equation}
The Lippmann--Schwinger equation now reads 
\begin{align}
\Psi_{\mathbf{\mathbf{k}}}(\mathbf{r}) & = & \phi_{\mathbf{\mathbf{k}}}(\mathbf{r})-\frac{ik}{4}\int d^{2}\mathbf{r}^{\prime}\left[H_{0}^{(1)}(k|\mathbf{r}-\mathbf{r}^{\prime}|)+\right.\nonumber \\
 &  & \left.i\sigma_{\theta}H_{1}^{(1)}(k|\mathbf{r}-\mathbf{r}^{\prime}|)\right]V(\mathbf{r}^{\prime})\Psi_{\mathbf{k}}(\mathbf{r^{\prime}})\,.\label{eq:scattered_wave_exact}
\end{align}
To proceed, we assume that the main contribution to the scattering
amplitude comes from evaluating the above integral within the region
where $|\mathbf{r}-\mathbf{r}^{\prime}|\gg1$. We note that although
this procedure is accurate for short-range potentials, yielding the
exact asymptotic form of the scattered wave function, it is otherwise
an approximation.

The next step is to insert the asymptotic expressions for the Hankel
functions 
\begin{equation}
H_{0}^{(1)}(k|\mathbf{r}-\mathbf{r}^{\prime}|)\rightarrow\sqrt{\frac{2}{ik\pi|\mathbf{r}-\mathbf{\mathbf{r}^{\prime}}|}}e^{ik|\mathbf{r}-\mathbf{r}^{\prime}|}\,,\label{eq:Hankel0_asymp}
\end{equation}
\begin{equation}
H_{1}^{(1)}(k|\mathbf{r}-\mathbf{r}^{\prime}|)\rightarrow-i\sqrt{\frac{2}{ik\pi|\mathbf{r}-\mathbf{\mathbf{r}^{\prime}}|}}e^{ik|\mathbf{r}-\mathbf{r}^{\prime}|},\label{eq:Hankel1_asymp}
\end{equation}
into the Lippmann--Schwinger equation (\ref{eq:scattered_wave_exact})
to get 
\begin{align}
\Psi_{\mathbf{\mathbf{k}}}(\mathbf{r}) & =\phi_{\mathbf{\mathbf{k}}}(\mathbf{r})-\sqrt{\frac{ik}{8\pi r}}e^{ikr}\int d^{2}\mathbf{r}^{\prime}e^{-i\mathbf{k^{\prime}}\cdot\mathbf{r}^{\prime}}\times\nonumber \\
 & \left(1+\sigma_{\theta}\right)V(\mathbf{r}^{\prime})\Psi_{\mathbf{k}}(\mathbf{r^{\prime}}).\label{eq:LS_final_form}
\end{align}
In the above, we have identified the wavevector at the point of observation,
$k^{\prime}\equiv k\frac{\mathbf{r}}{r},$ and used $|\mathbf{r}-\mathbf{r}^{\prime}|\simeq\mathbf{r}-\mathbf{r}\cdot\mathbf{r}^{\prime}/r$
to simplify the argument of the exponentials in (\ref{eq:Hankel0_asymp})--(\ref{eq:Hankel1_asymp}).

The first term in the Born series is obtained by replacing $\Psi_{\mathbf{k}}(\mathbf{r^{\prime}})\rightarrow\phi_{\mathbf{k}}(\mathbf{r^{\prime}})=e^{i\mathbf{k}\cdot\mathbf{r}^{\prime}}u_{\mathbf{k}}$
in the right-hand side of the Lippmann--Schwinger equation. In order
to read out the scattering amplitude a few manipulations are still
in order. Without loss of generality, setting $\theta_{\mathbf{k}}=0$,
and identifying $\theta_{\mathbf{k}^{\prime}}$ with the scattering
angle $\theta$, we find 
\begin{eqnarray}
\left(1+\sigma_{\theta}\right)u_{\mathbf{k}} & = & (1+e^{-i\theta})\frac{1}{\sqrt{2}}\left(\begin{array}{c}
1\\
e^{i\theta}
\end{array}\right)\label{eq:map1}\\
 & \equiv & \Xi(\theta)u_{\mathbf{k}^{\prime}},\label{eq:map2}
\end{eqnarray}
where we have defined graphene's Berry phase (form factor) term $\Xi(\theta)=1+e^{-i\theta}$.
Substituting this result into Eq.~(\ref{eq:LS_final_form}) we arrive
at the well-known FBA in two-dimensions 
\begin{equation}
\Psi_{\mathbf{k}}(\mathbf{r})=\phi_{\mathbf{\mathbf{k}}}(\mathbf{r})+\frac{f_{\textrm{FBA}}(\theta)}{\sqrt{r}}e^{ikr}u_{\mathbf{k^{\prime}}}.\label{eq:scattered_wave_FBA}
\end{equation}
with 
\begin{equation}
f_{\textrm{FBA}}(\theta)=-\frac{1}{v_{F}\hbar}\sqrt{\frac{ik}{8\pi}}\Xi(\theta)\tilde{V}(\mathbf{q}),\label{eq:f_FBA}
\end{equation}
where $\mathbf{q}=\mathbf{k^{\prime}-}\mathbf{k}$ is the transferred
wavevector and the relevant constants have been restored. We note
that the above definition yields the usual form for the scattered
current in two-dimensions, i.e., 
\begin{equation}
J(\theta)=\langle\tilde{\Psi}_{\mathbf{k}}(\mathbf{r})|\sigma_{\theta}|\tilde{\Psi}_{\mathbf{k}}(\mathbf{r})\rangle\propto\frac{|f_{\textrm{FBA}}(\theta)|^{2}}{r}\,,\label{eq:scattered_current}
\end{equation}
with $\tilde{\Psi}_{\mathbf{k}}(\mathbf{r})\equiv\Psi_{\mathbf{k}}(\mathbf{r})-\phi_{\mathbf{\mathbf{k}}}(\mathbf{r})$
denoting the scattered component of the wave.

We move gears to the calculation of the second term in the Born series.
The starting point is Eq.~(\ref{eq:LS_position_rep}), which we iterate
two times to get 
\begin{eqnarray}
\Psi_{\mathbf{k}}(\mathbf{r})=\phi_{\mathbf{k}}(\mathbf{r})+\int d^{2}\mathbf{r}^{\prime}\left[G_{0}(\mathbf{r}-\mathbf{r}^{\prime})V(\mathbf{r}^{\prime})\phi_{\mathbf{k}}(\mathbf{r^{\prime}})+\right.\nonumber \\
\left.\int d^{2}\mathbf{r}^{\prime\prime}G_{0}(\mathbf{r}-\mathbf{r}^{\prime})V(\mathbf{r}^{\prime})G_{0}(\mathbf{r}^{\prime}-\mathbf{r}^{\prime\prime})V(\mathbf{r}^{\prime\prime})\phi_{\mathbf{k}}(\mathbf{r^{\prime\prime}})\right]\,.\label{eq:Lipp_2times}
\end{eqnarray}
We aim to simplify the second order contribution in the above expression
{[}from now on referred to as $\Psi_{\mathbf{k}}^{(2)}(\mathbf{r})${]}.
As before, we replace $G_{0}(\mathbf{r}-\mathbf{r}^{\prime})$ by
its asymptotic form

\begin{equation}
G_{0}(\mathbf{r}-\mathbf{r}^{\prime})\rightarrow-\sqrt{\frac{ik}{8\pi r}}e^{ikr}e^{-i\mathbf{k^{\prime}}\cdot\mathbf{r}^{\prime}}\tilde{\sigma}_{\theta},\label{eq:A4-6-21}
\end{equation}
with $\tilde{\sigma}_{\theta}\equiv1+\sigma_{\theta}$, and insert
it back into $\Psi_{\mathbf{k}}^{(2)}(\mathbf{r})$ as to obtain 
\begin{equation}
\Psi_{\mathbf{k}}^{(2)}(\mathbf{r})=-\sqrt{\frac{ik}{8\pi r}}e^{ikr}\Upsilon_{\mathbf{k}\mathbf{k}^{\prime}}\,,\label{eq:palmeira}
\end{equation}
where 
\begin{equation}
\Upsilon_{\mathbf{k}\mathbf{k}^{\prime}}=\int d^{2}\mathbf{r}^{\prime}d^{2}\mathbf{r}^{\prime\prime}e^{-i\mathbf{k^{\prime}}\cdot\mathbf{r}^{\prime}}\tilde{\sigma}_{\theta}V(\mathbf{r}^{\prime})G_{0}(\mathbf{r}^{\prime}-\mathbf{r}^{\prime\prime})V(\mathbf{r}^{\prime\prime})\phi_{\mathbf{k}}(\mathbf{r^{\prime\prime}})\,.\label{eq:palmeira_exp}
\end{equation}
It is clear that $\tilde{\sigma}_{\theta}$ does not commute with
the remaining terms in the integrand {[}remark that $G_{0}(\mathbf{r}^{\prime}-\mathbf{r}^{\prime\prime})$
contains a term proportional to $\sigma_{\alpha}$ with $\alpha\equiv\theta(\mathbf{r}^{\prime},\mathbf{r}^{\prime\prime})\neq\theta$;
c.f.,~Eq.(\ref{eq:GF_position}){]}, and hence we cannot directly
identify the scattering amplitude as previously. Instead, we make
use of Eq.~(\ref{eq:GF_1}) to write 
\begin{eqnarray}
\Upsilon_{\mathbf{k}\mathbf{k}^{\prime}}=\int\frac{d^{2}\mathbf{p}}{\left(2\pi\right)^{2}}g(\mathbf{p})\int d^{2}\mathbf{r}^{\prime\prime}V(\mathbf{r}^{\prime\prime})e^{-i(\mathbf{p}-\mathbf{k})\cdot\mathbf{r}^{\prime\prime}}\times\nonumber \\
\left[\int d^{2}\mathbf{r}^{\prime}e^{-i\mathbf{k^{\prime}}\cdot\mathbf{r}^{\prime}}V(\mathbf{r}^{\prime})\tilde{\sigma}_{\theta}\left(k-i\sigma\cdot\boldsymbol{\nabla}^{\prime}\right)e^{i\mathbf{p}\cdot\mathbf{r}^{\prime}}\right]u_{\mathbf{k}}\,,\label{eq:palmeira_simp}
\end{eqnarray}
or, using the definition of Fourier transform, 
\begin{eqnarray}
\Upsilon_{\mathbf{k}\mathbf{k}^{\prime}}=\int\frac{d^{2}\mathbf{p}}{\left(2\pi\right)^{2}}\tilde{V}(\mathbf{p}-\mathbf{k})\left[\tilde{\sigma}_{\theta}G_{0}(\mathbf{p})\right]\tilde{V}(\mathbf{k}^{\prime}-\mathbf{p})u_{\mathbf{k}}\,.\label{eq:palmeira_simp_2}
\end{eqnarray}
In order to identify the scattering amplitude in the second Born approximation
(SBA) we compute the contribution of $\Psi_{\mathbf{k}}^{(2)}(\mathbf{r})$
to the scattering flux. Neglecting terms of fourth order in the scattering
potential, we find 
\[
J_{\textrm{SBA}}(\theta)=\langle\tilde{\Psi}_{\mathbf{k}}(\mathbf{r})|\sigma_{\theta}|\tilde{\Psi}_{\mathbf{k}}(\mathbf{r})\rangle=J(\theta)+\delta J(\theta)\,,
\]
with $J(\theta)$ given by Eq.~(\ref{eq:scattered_current}) and
\begin{equation}
\delta J(\theta)=-\frac{f_{\textrm{FBA}}^{*}(\theta)}{r}\sqrt{\frac{ik}{8\pi}}\langle u_{\mathbf{k^{\prime}}}|\sigma_{\theta}|\Upsilon_{\mathbf{k}\mathbf{k}^{\prime}}\rangle+\mathcal{\textrm{c.c.}}\:.\label{eq:delta_J}
\end{equation}
Using 
\begin{eqnarray}
\sqrt{2}\tilde{\sigma}_{\theta}\left(k+\sigma\cdot\mathbf{p}\right)u_{\mathbf{k}}=\left(\begin{array}{c}
\Xi(\theta)k+pe^{-i\phi_{\mathbf{p}}}+pe^{i(\phi_{\mathbf{p}}-\theta)}\\
\Xi(-\theta)k+pe^{i\phi_{\mathbf{p}}}+pe^{i(\theta-\phi_{\mathbf{p}})}
\end{array}\right)\,,\label{eq:aux_iden}
\end{eqnarray}
where $\phi_{\mathbf{p}}=\arctan(p_{y}/p_{x})$, we arrive at the
following result 
\begin{align}
\langle u_{\mathbf{k^{\prime}}}|\sigma_{\theta}|\Upsilon_{\mathbf{k}\mathbf{k}^{\prime}}\rangle & =\int\frac{d^{2}\mathbf{p}}{\left(2\pi\right)^{2}}\tilde{V}(\mathbf{p}-\mathbf{k})\left[k+pe^{-i\phi}+\right.\nonumber \\
 & \left.+pe^{i(\phi-\theta)}+ke^{-i\theta}\right]\tilde{V}(\mathbf{k}^{\prime}-\mathbf{p})\,.\label{eq:aux_res}
\end{align}
By the definition of scattered current $J_{\textrm{SBA}}(\theta)$,
the SBA scattering amplitude is readily seen to be 
\begin{align}
f_{\textrm{SBA}}(\theta) & =\sqrt{\frac{k}{8\pi}}\left\{ \Xi(\theta)\tilde{V}(\mathbf{k}^{\prime}-\mathbf{k})+\right.\nonumber \\
 & \int\frac{d^{2}\mathbf{p}}{\left(2\pi\right)^{2}}\tilde{V}(\mathbf{k}^{\prime}-\mathbf{p})\left[\Xi(\theta)\left(k+p\cos\phi_{\mathbf{p}}\right)+\right.\nonumber \\
 & \left.\left.\bar{\Xi}(\theta)ip\sin\phi_{\mathbf{p}}g(\mathbf{p})\right]\tilde{V}(\mathbf{p}-\mathbf{k})\right\} \,,\label{eq:f_SBA}
\end{align}
where we defined $\bar{\Xi}(\theta)=\Xi(\theta+\pi)$ and dropped
an innocuous phase factor $-\sqrt{i}$. We now specialize to potentials
with inversion symmetry; these potentials have $\tilde{V}(\mathbf{q})=\tilde{V}(\mathbf{q})^{*}$
and therefore we can drop the imaginary term in last line of Eq.~(\ref{eq:f_SBA}),
which is odd under the transformation $\theta\rightarrow-\theta$,
and hence does not contribute to transport cross sections. We thus
arrive at our desired result 
\begin{align}
f_{\textrm{SBA}}(\theta) & =\Xi(\theta)\sqrt{\frac{k}{8\pi}}\left\{ \tilde{V}(\mathbf{k}^{\prime}-\mathbf{k})+\int\frac{d^{2}\mathbf{p}}{\left(2\pi\right)^{2}}\times\right.\label{eq:f_SBA_2}\\
 & \left.\tilde{V}(\mathbf{k}^{\prime}-\mathbf{p})\left(k+p\cos\phi_{\mathbf{p}}\right)g(\mathbf{p})\tilde{V}(\mathbf{p}-\mathbf{k})\right\} \,,\nonumber 
\end{align}
or, in a more compact form, 
\begin{eqnarray}
f_{\textrm{SBA}}(\theta) & = & \frac{\Xi(\theta)}{v_{F}\hbar}\sqrt{\frac{k}{8\pi}}\left[\tilde{V}(\mathbf{k}^{\prime}-\mathbf{k})+\int\frac{d^{2}\mathbf{p}}{\left(2\pi\right)^{2}}\times\right.\nonumber \\
 &  & \left.\tilde{V}(\mathbf{k}^{\prime}-\mathbf{p})\langle u_{\mathbf{k}}|G_{0}(\mathbf{p})|u_{\mathbf{k}}\rangle\tilde{V}(\mathbf{p}-\mathbf{k})\right],\label{eq:f_SBA_MAIN}
\end{eqnarray}
where $\hbar$ and $v_{F}$ have been restored.

\section*{Appendix D: Calculation of second Born amplitude for a charged line}

In this appendix we evaluate the SBA transport cross section for a
charged line with potential given by Eq.~(\ref{eq:potential_finite_line}).
We perform an analytical calculation of the $\mathcal{I}_{1}$ contribution
{[}Eq.~(\ref{eq:a1_SBA}){]} and evaluate the remaining contribution
{[}Eq.~(\ref{eq:a2_SBA}){]} numerically. The term $\mathcal{I}_{1}$
requires to evaluate the following integral 
\begin{align}
\eta_{1} & =\int\frac{d^{2}\mathbf{p}}{\left(2\pi\right)^{2}}\frac{1}{k^{2}-\mathbf{p}^{2}+i0^{+}}\nonumber \\
 & \qquad\frac{1}{q_{\mathtt{TF}}+|p_{x}-k_{x}|}\frac{1}{q_{\mathtt{TF}}+|k_{x}^{\prime}-p_{x}|},\label{eq:eta_1}
\end{align}
which we do by first performing the integration over $k_{y}$ to get
\begin{eqnarray}
\eta_{1} & = & \int\frac{dp_{x}}{2\pi}\frac{i}{2p_{0}}\frac{1}{q_{\mathtt{TF}}+|p_{x}-k_{x}|}\frac{1}{q_{\mathtt{TF}}+|k_{x}^{\prime}-p_{x}|}\,,\label{eq:Re_Eta_1_b}
\end{eqnarray}
where $p_{0}=\sqrt{k^{2}-p_{x}^{2}}+i0^{+}$. To proceed, we divide
the integration range into four subintervals: $p_{x}\ge k_{x}$, $k_{x}>p_{x}\ge k_{x}^{\prime}$,
$k_{x}^{\prime}>p_{x}\ge-k_{x}$ and $p_{x}<-k_{x}$. Each of these
contributions has a solution in terms of simple functions. We give
the explicit solution for the real part of $\eta_{1}$. Since $p_{0}$
becomes pure imaginary for $|p_{x}|\ge k$, we have 
\begin{align}
\textrm{Re}\,\eta_{1} & =\frac{1}{2}\left\{ \int_{-\infty}^{-k}\frac{dp_{x}}{2\pi}\frac{1}{\sqrt{p_{x}^{2}-k^{2}}}\times\right.\nonumber \\
 & \qquad\frac{1}{q_{\mathtt{TF}}-p_{x}+k_{x}}\frac{1}{q_{\mathtt{TF}}+k_{x}^{\prime}-p_{x}}+\nonumber \\
 & \qquad\int_{k}^{\infty}\frac{dp_{x}}{2\pi}\frac{1}{\sqrt{p_{x}^{2}-k^{2}}}\times\nonumber \\
 & \left.\qquad\frac{1}{q_{\mathtt{TF}}+p_{x}-k_{x}}\frac{1}{q_{\mathtt{TF}}-k_{x}^{\prime}+p_{x}}\right\} .\label{eq:Re_eta_1}
\end{align}
Without loss of generality we set $k_{x}=k$, $k_{x}^{\prime}=k\cos\theta$.
The integral above then acquires the form 
\begin{equation}
\textrm{Re}\,\eta_{1}=\frac{1}{32\pi\alpha_{\textrm{g}}}\frac{\chi(\theta)}{k_{F}^{2}\left(1-\cos\theta\right)},\label{eq:Re_Eta_1_d}
\end{equation}
with $k=k_{F}$ and 
\begin{align}
\chi(\theta) & =\frac{-2\arccos\left(1+4\alpha_{\textrm{g}}\right)}{i\sqrt{1+\frac{1}{2\alpha_{\textrm{g}}}}}+\frac{\pi+2\arcsin\left(1-4\alpha_{\textrm{g}}\right)}{\sqrt{-1+\frac{1}{2\alpha_{\textrm{g}}}}}-\nonumber \\
 & \frac{8\alpha_{\textrm{g}}\arccos\left(4\alpha_{\textrm{g}}-\cos\theta\right)}{\sqrt{1-\left(4\alpha_{\textrm{g}}-\cos\theta\right)^{2}}}+\frac{8\alpha_{\textrm{g}}\textrm{arccosh}\left(4\alpha_{\textrm{g}}+\cos\theta\right)}{\sqrt{\left(4\alpha_{\textrm{g}}+\cos\theta\right)^{2}-1}}\,,\label{eq:chi}
\end{align}
and where used $q_{\mathtt{TF}}=4\alpha_{\textrm{g}}k_{F}$. The remaining
term to be computed reads 
\begin{align}
\eta_{2} & =\int\frac{d^{2}\mathbf{p}}{\left(2\pi\right)^{2}}\frac{p_{x}}{k^{2}-\mathbf{p}^{2}+i0^{+}}\nonumber \\
 & \qquad\frac{1}{q_{\mathtt{TF}}+|k_{x}^{\prime}-p_{x}|}\frac{1}{q_{\mathtt{TF}}+|q_{x}-k_{x}|}\,.\label{eq:eta_2}
\end{align}
The explicit form of $\eta_{2}$ is rather cumbersome and thus will
not be given. The differential cross section is 
\begin{equation}
\sigma(\theta)=|f_{1}(\theta)+f_{2}(\theta)|^{2}\,,\label{eq:SBA_diff_cross}
\end{equation}
where $f_{1(2)}$ denotes the first (second) order contribution to
the SBA amplitude {[}see Eq.~(\ref{eq:f_SBA_MAIN}){]}. Defining
$h_{1(2)}(\theta)\equiv f_{1(2)}(\theta)/\Xi(\theta)$, we obtain
\begin{equation}
\sigma(\theta)=|\Xi(\theta)|^{2}\left\{ h_{1}(\theta)^{2}+2\textrm{Re}\left[h_{2}(\theta)\right]h_{1}(\theta)+\mathcal{O}(\Delta^{4})\right\} ,\label{eq:SBA_diff_cross_sec}
\end{equation}
and where we have used the fact that $h_{1}(\theta)\in\mathbb{R}$
for potentials with inversion symmetry. The first term yields the
FBA transport cross section 
\begin{eqnarray}
\sigma_{\textrm{tp}}^{(\textrm{FBA})} & = & \int_{0}^{2\pi}d\theta\left(1-\cos\theta\right)|f_{1}(\theta)|^{2}\nonumber \\
 & = & \frac{k_{F}}{8\pi}\left(\frac{L\Delta}{\hbar v_{F}}\right)^{2}\int_{0}^{2\pi}d\theta\frac{\left(1-\cos\theta\right)|\Xi(\theta)|^{2}}{\left[2k_{F}\sin^{2}(\theta/2)+q_{\mathtt{TF}}\right]^{2}}\nonumber \\
 & = & \left(\frac{L\Delta}{\hbar v_{F}}\right)^{2}\frac{Q_{1}(\alpha)}{k_{F}},\label{eq:I1}
\end{eqnarray}
with 
\begin{equation}
Q_{1}(\alpha_{\textrm{g}})\equiv\frac{1}{8\pi}\int_{0}^{2\pi}d\theta\frac{\left(1-\cos\theta\right)|\Xi(\theta)|^{2}}{\left[2\sin^{2}(\theta/2)+4\alpha_{\textrm{g}}\right]^{2}}\,.\label{eq:I1_exp}
\end{equation}
Remark that the transport relaxation rate is related to $\sigma_{\textrm{tp}}$
according to $\tau=(n_{l}v_{F}\sigma_{\textrm{tp}})^{-1}\sim k_{F}$,
and therefore we conclude that the dc-conductivity 
\begin{equation}
\sigma=\frac{2e^{2}}{h}v_{F}k_{F}\tau(k_{F})=\frac{2e^{2}}{h}\frac{k_{F}}{n_{l}\sigma_{\textrm{tp}}(k_{F})}\,,\label{eq:relation_sigmadc_sigmatp}
\end{equation}
is a quadratic (linear) function of the Fermi wavevector (electronic
density). The latter property is preserved at all orders in perturbation
theory as noted in Sec.~\ref{sec:Boltzmann_Approach}.

The second term in Eq.~(\ref{eq:SBA_diff_cross_sec}) yields the
main correction to the FBA transport cross section; explicitly, 
\begin{align}
\delta\sigma_{\textrm{tp}}= & \frac{k_{F}}{8\pi}\left(\frac{L\Delta}{\hbar v_{F}}\right)^{3}2k_{F}\int_{0}^{2\pi}d\theta\left(1-\cos\theta\right)\times\nonumber \\
 & \qquad|\Xi(\theta)|^{2}\frac{\textrm{Re}\left[\eta_{1}(\theta)+\eta_{2}(\theta)\right]}{2k_{F}\sin^{2}(\theta/2)+q_{\mathtt{TF}}}\,.\label{eq:delta_sigma_tp}
\end{align}
Simplifying one obtains 
\begin{equation}
\delta\sigma_{\textrm{tp}}=\left(\frac{L\Delta}{\hbar v_{F}}\right)^{3}\frac{Q_{2}(\alpha_{\textrm{g}})}{k_{F}}\,,\label{eq:delta_sigma_t_2}
\end{equation}
with 
\begin{equation}
Q_{2}(\alpha_{\textrm{g}})\equiv\frac{1}{8\pi}\int_{0}^{2\pi}d\theta\frac{\left(1-\cos\theta\right)|\Xi(\theta)|^{2}}{2\sin^{2}(\theta/2)+4\alpha_{\textrm{g}}}\psi_{\textrm{SBA}}(\theta)\,,\label{eq:I2}
\end{equation}
and $\psi_{\textrm{SBA}}(\theta)\equiv2k_{F}^{2}\textrm{Re}\left[\eta_{1}(\theta)+\eta_{2}(\theta)\right]$
is just a function of $\theta$ and $\alpha_{\textrm{g}}$ {[}recall
that $\eta_{1(2)}$ varies as $k_{F}^{-2}$; see Eq.~(\ref{eq:Re_Eta_1_d}){]}.
Finally, one obtains for the SBA transport cross section 
\begin{eqnarray}
\sigma_{\textrm{tp}}^{(\textrm{SBA})} & = & \sum_{n=2,3}\left(\frac{L\Delta}{\hbar v_{F}}\right)^{n}\frac{Q_{n-1}(\alpha_{\textrm{g}})}{k_{F}}+\mathcal{O}(\Delta^{4}).\label{eq:sigma_tp_SBA}
\end{eqnarray}
Collecting these results one obtains the following relation between
the SBA and the FBA conductivities 
\begin{equation}
\frac{\sigma_{\textrm{dc}}^{(\textrm{SBA})}}{\sigma_{\textrm{dc}}^{(\textrm{FBA})}}=\frac{\sigma_{\textrm{tp}}^{(\textrm{FBA})}}{\sigma_{\textrm{tp}}^{(\textrm{SBA})}}=1-\frac{L\Delta}{\hbar v_{F}}\frac{Q_{2}(\alpha_{\textrm{g}})}{Q_{1}(\alpha_{\textrm{g}})}+\mathcal{O}\left(\frac{L\Delta}{\hbar v_{F}}\right)^{2}.\label{eq:sigma_SBA_VS_sigma_FBA}
\end{equation}
The above result shows that for $\Delta>0$ ($\Delta<0$) the SBA
decreases (increases) the dc conductivity with respect to the FBA
result. Although only valid in the weak scattering regime, this dependence
of the dc conductivity on the carrier polarity is in qualitatively
agreement with the numerical results of Sec.~\ref{sec:TB_Kubo_RealSpace}.


\begin{thebibliography}{10}
\bibitem{comment}The numerical evaluation of the Kubo formula in
the weak scattering requires extremely large computational domains
(needed to reach a diffusive transport) which is beyond our computational
capabilities.
\bibitem{finite_line}The potential energy in Fourier space $V_{L}(\boldsymbol{q})$
considered here is obtained from a toy model of a finite line with
$V_{L}(x,y)=\pm e\phi(x)g(y)$, with $\phi(x)$ being the electrostatic
potential of an infinite line {[}Eq.\,(\ref{eq:effective_potential_real_space}){]},
and $g(y)$ a smooth function of $y$ with the following properties:
(1) $g(y)$ is symmetric under reflection, and (2) $g(y)$ varies
slowly for $|y|\ll l$ and vanishes exponentially as $|y|\gg l$,
where $l\lesssim L$. For instance, choosing $g(y)=(L/4\pi)e^{-|y|/l}$,
we obtain $\tilde{V}_{L}(\boldsymbol{q})=(L/2\pi)V(q_{x})l^{-1}(q_{y}^{2}+l^{-2})^{-1}$
which upon substitution of $V(q_{x})$ and taking $q_{y}l\ll1$ yields
$\tilde{V}_{L}(\boldsymbol{q})\simeq L\Delta/(|q_{x}|+q_{\textrm{TF}})$.
We note that the same result can be obtained by broadening the delta
function $\delta(q_{y})$ appearing in the potential of an infinite
line.
\end{thebibliography}
\end{document}